\documentclass[12pt]{article}
\usepackage[a4paper,margin=2.5cm]{geometry}
\usepackage[T1]{fontenc}
\usepackage[utf8]{inputenc}
\usepackage[natbibapa]{apacite}
\usepackage[utf8]{inputenc}
\usepackage{IEEEtrantools}
\usepackage[english]{babel}
\usepackage{amsmath,amsthm,amssymb}
\usepackage{graphicx}
\usepackage{bm}
\usepackage[flushleft]{threeparttable}
\usepackage{colortbl}
\usepackage{graphicx}
\usepackage{pdflscape}
\usepackage{caption}
\usepackage{subcaption}
\usepackage{enumitem}
\usepackage{graphicx}
\usepackage{bm}
\usepackage{textcomp}
\usepackage{textgreek}
\usepackage{multicol}
\usepackage{setspace}
\usepackage{IEEEtrantools}
\usepackage{booktabs}
\usepackage{authblk}
\usepackage{float}

\usepackage{mathtools}

\usepackage{setspace}
\usepackage[final]{changes}
\usepackage[colorinlistoftodos,prependcaption]{todonotes}

\usepackage{titlesec}
\titleformat{\paragraph}
{\normalfont\normalsize\bfseries}{\theparagraph}{1em}{}
\titlespacing*{\paragraph}
{0pt}{3.25ex plus 1ex minus .2ex}{1.5ex plus .2ex}

\DeclareMathOperator{\Cov}{\mathbb{C}ov}

\title{A Square-Root Second-Order Extended Kalman Filtering Approach for Estimating Smoothly Time-Varying Parameters}

\date{}

\author{
Zachary F. Fisher \\
\vspace{-.4cm}
Sy-Miin Chow \\
Peter C. M. Molenaar \\
Barbara L. Fredrickson \\
Vladas Pipiras \\
Kathleen M. Gates}

\begin{document}
\makeatletter
\@ifundefined{@affil}{\def\@affil{~}{}}
\makeatother
\maketitle

\newpage
\doublespacing

\begin{abstract}
Researchers collecting intensive longitudinal data (ILD) are increasingly looking to model psychological processes, such as emotional dynamics, that organize and adapt across time in complex and meaningful ways. This is also the case for researchers looking to characterize the impact of an intervention on individual behavior. To be useful, statistical models must be capable of characterizing these processes as complex, time-dependent phenomenon, otherwise only a fraction of the system dynamics will be recovered. In this paper we introduce a Square-Root Second-Order Extended Kalman Filtering approach for estimating smoothly time-varying parameters. This approach is capable of handling dynamic factor models where the relations between variables underlying the processes of interest change in a manner that may be difficult to specify in advance.  We examine the performance of our approach in a Monte Carlo simulation and show the proposed algorithm accurately recovers the unobserved states in the case of a bivariate dynamic factor model with time-varying dynamics and treatment effects. Furthermore, we illustrate the utility of our approach in characterizing the time-varying effect of a meditation intervention on day-to-day emotional experiences. 
\end{abstract}

\section{Introduction}

Reality is complicated. This is especially true in the psychological sciences where the modeling of basic psychological processes must contend with large amounts of measurement error, nonlinear relations among phenomena of interest and often severe unobserved heterogeneity. In terms of heterogeneity, not only do individuals differ from one another in complex and meaningful ways, psychological processes within individuals develop and adapt across a myriad of timescales and contexts.  To be useful, models must be capable of characterizing these processes as complex, time-dependent phenomenon, otherwise the resulting insights and decisional criteria afforded by the modeling process will be narrowly defined. In this paper we introduce a nonlinear state space model and estimation framework capable of handling complex dynamic models where the relations between variables underlying processes of interest change in a manner that may be difficult to specify in advance. That is to say we outline a method that is useful even when there is little pre-existing knowledge about the nature of the change process itself. In the remainder of this introduction we will motivate the proposed model and estimator for psychological researchers by providing the requisite background information on our implementation while explicitly detailing the types of processes and questions amenable to this modeling framework.

\subsection{Nonstationary Processes in Psychological Research}

Historically, the majority of the probability theory developed for time series  analysis was concerned with stationary time series. To define a stationary process let $\mathbf{y}_{t}$ be a $k\times1$ vector of observations at time $t$.  We can call the process $\mathbf{y}_{t}$ stationary if the probability distributions from the random vectors $(\mathbf{y}_{t_1},\dots,\mathbf{y}_{t_n})$ and $(\mathbf{y}_{{t_1}+\ell},\dots,\mathbf{y}_{t_{n}+\ell})$ are equivalent for all lags or leads ($\ell=0,\pm1,\pm2,\dots$) and all set of times ($t_{1},\dots. t_{n}$) \citep[p. 29-33]{grenander1957}. Intuitively this means a time series is stationary if there are no systematic changes in the series mean (e.g. trends) or variance, and all periodic variation has been removed from the series. Technically speaking, the majority of natural processes are unlikely to be stationary but it is also often the case that nonstationary series can be made stationary for the purpose of analysis.  As the lion's share of analytic approaches assume  stationarity this is a convenient choice for researchers when a more complex characterization of the system is not possible.  

However, unlike many time series modelers  who are primarily interested in forecasting, psychologists are generally focused on model construction and interpretation, and removing nonstationary characteristics of the process is in many cases inconsistent with the goals of the modeling endeavor. If researchers are unable to approximate the complexity of the process under study, the model itself is unlikely to provide useful insights about the phenomenon. Furthermore, failure to account for nonstationarity can lead to dramatic underestimates of the uncertainty associated with a given model, leading researchers to be overconfident in their assessments, making generalization and the establishment of lawful relations based on individual behavior more difficult.

For these reasons models of psychological processes developed from sequentially collected experimental or observational data often require methodological approaches capable of handling nonstationarity. For example, if the subject under study adjusts their responses based on changing decisional criteria, concentration levels, fatigue or any factors secondary to the stimuli itself the process is likely to be nonstationary. Examples of this can be found in models of learning \citep{boker2007}, psychophysical stimulus detection paradigms where responses may depend on more than just the task intensity at any given presentation \citep{doll2015,frund2011}, as well as emotional dynamics both within \citep{koval2012} and between \citep{bringmann2018} individuals, to name just a few. Furthermore, by design the presence of an intervention will often lead a process to become nonstationary.  This is even more likely if the impact of the intervention changes throughout (and possibly following) the course of a treatment. Related to the examples given above but stated more generally stationarity implicitly requires the parameters or relations among variables underlying a phenomenon of interest to be invariant with respect to time and violation of this premise will lead to a nonstationary series. 

\subsection{Time-Varying Parameter Models}

Models that allow for parameters to change across time are one approach for handling nonstationarity. Broadly these models can be classified as Regime-Switching (RS) when parameters are allowed to vary discretely, typically as a function of recurrent shifts between distinct model states (or regimes).  A second class of models, and the models with which this paper is concerned, are time-varying parameter (TVP) models where the parameters are allowed to continuously vary across time. In the TVP case changes are hypothesized to be smooth rather than sharp. In the case of observed variable time series popular RS models include the threshold autoregressive model \citep{tong1980},  threshold cointegration model \citep{balke1997}, the threshold unit root model \citep{caner2001}, and the Markov-switching autoregressive model \citep{hamilton1989}. Psychological researchers have employed these models to investigate dyadic interactions \citep{hamaker2009}, the time dependency between positive and negative affect \citep{hamaker2010}, and have also extended these frameworks to handle time series from multiple individuals \citep{dehaan2016}.  

Likewise, commonly implemented TVP models for observed variable time series include the the local linear trend model \citep{harvey1990}, time varying autoregressive moving average (ARMA) models (where the autoregression parameter is allowed to vary over time) \citep{weiss1985}, the stochastically varying coefficient regression (VCR) model \citep{pagan1980} and the time-varying vector autoregression (VAR) model \citep{jiang1993}. TVP models have also been employed in psychological applications.  For example, time-varying VAR models have been used to investigate emotion dynamics among dyads \citep{bringmann2018} and well-being in individuals diagnosed with Autism Spectrum Disorder \citep{haslbeck2019}. Interested readers can see \citet{haslbeck2019b} for a detailed overview and an empirical comparison of the approaches developed by \citet{bringmann2018} and \citet{haslbeck2019}.

Unfortunately many of the processes most interesting to psychologists are complex, characterized by nonlinear relations among variables and diluted by measurement error.  When unaccounted for measurement error can cause unintended consequences during modeling for both time-invariant and time-varying parameter models. None of the observed variable approaches described above are intended to account for measurement error.  When present and unaccounted for measurement error will render an otherwise observed AR process latent, and this latent process will be of a differing order than the observed process \citep{box1976}, complicating the modeling process.  Furthermore, if an AR model is fit to observed data measured with error the autoregressive coefficients obtained from this analysis will be biased towards zero \citep{staudenmayer2005} and maximum likelihood estimation will provide unreliable inferences \citep{fuller2009} and distorted diagnostic tests \citep{patriota2010}. For these reasons researchers will often turn to state space (or factor analytic) methods when multiple-indicators of a construct of interest are available. State space methods are capable of accounting for measurement error while elegantly handling a wide variety of nonlinear dependencies among latent and observed variables. Similar to the observed variable case, this class of nonlinear latent variable models with time-varying parameters have also been employed by psyshometricians and researchers in the social and behavioral sciences.

In the economics literature \citet{stock2009} proposed an exploratory dynamic factor model to identify discontinuities within economic time series evidenced by discrete shifts in the factor loading pattern.  \citet{yang2010} proposed a regime-switching approach within the state-space modeling framework for characterizing the dynamics of facial electromyography. Here the discrete shifts alone rendered the process nonlinear, however, the latent states themselves were characterized by a linear process within each regime. \citet{chow2013} extended the model from \citet{yang2010} to the nonlinear case in the form of a nonlinear regime-switching state-space model estimated using a combination of the Extended Kalman and Kim Filters. Here the nonlinearity arises from the dynamics within each regime being defined according to the logistic function and the piecewise nonlinearity introduced by the regime switching. Although the class of hidden Markov models are similar to the latent variable RS models described above they are generally defined without dynamics among the latent states and for this reason we do not consider them further here. 

In the psychology literature \citet{molenaar1992} was the first to propose a unidimensional dynamic factor model for nonstationary data by incorporating a linear trend at the latent level.   \citet{molenaar1994} extended this model further to allow for the autoregressive and factor loading parameters to vary as polynomial functions of time.  In the economics literature \citet{negro2008} proposed a dynamic factor model with arbitrarily time-varying factor loadings and stochastic volatility in both the latent factors and error components.  \citet{chow2011} also developed a dynamical factor model with a single time-varying cross-regressive parameter hypothesized to obey an AR(1) process based on the differences in the coefficient values from baseline.  This model was estimated using a first-order Extended Kalman Filter (EKF), similar to the method proposed in \citet{molenaar1994}.  In the neuroimaging literature a number of authors have explored using the EKF to model physiological signals. For example, \citet{milde2010} used a version of the EKF to model high-dimensional multi-trial laser-evoked brain potentials. \citet{havlicek2011} applied the EKF to coupled dynamical systems of evoked brain responses in functional Magnetic Resonance Imaging (fMRI). \citet{hu2012} used the EKF
 to model time-varying source connectivity based on somatosensory evoked potentials.  Finally, \citet{molenaar2016} proposed a multi-dimensional exploratory factor analysis model in conjunction with a second-order Extended Kalman Filter (SEKF) to estimate individual-level functional connectivity maps from fMRI data. 

The current work directly extends the procedures developed by \citet{molenaar2016} in several important ways. First, we have developed a square-root version of the second-order Extended Kalman Filter (SR-SEKF) which effectively doubles the numerical precision of the SEKF algorithm. Second, we have implemented the Rauch-Tung-Striebel \citep[RTS;][]{rauch1965} smoother adapted to the second-order Extended Kalman Filter. Third, we examine the performance of the developed procedure in the context of detecting and estimating multiple time-varying parameters simultaneously. Fourth, we examine the performance of the algorithm in detecting and characterizing multiple time-varying treatment effects. Fifth, we systematically explore the impact of allowing for more time-varying parameters on the bias and variability of the existing parameter estimates.

The remainder of the article is organized as follows.  We begin by orienting the reader to the linear dynamic factor analysis model as typically presented in the structural equation modeling (SEM) framework.  We then demonstrate how the linear model can be adapted to handle time-varying parameters and detail a set of estimation routines that are capable of handling the nonlinearities induced by this adaptation.  We examine the performance of the estimator under a number of novel modeling conditions and provide an empirical example demonstrating the utility of the proposed approach for psychological researchers.

\section{Models and Notation}

\subsection{Linear Dynamic Factor Model Specification}

With minor modifications we use the notation of \citet{molenaar2017} to specify a general form for the dynamic factor model:

\begin{align}
\mathbf{y}_{t} &= 
  \sum^{s}_{u=0} 
  \boldsymbol{\Lambda}_{u}
  \boldsymbol{\eta}_{t-u} +
  \boldsymbol{\varepsilon}_{t}
  \label{ov.eqn}
  \\
\boldsymbol{\eta}_{t} &=
	\sum^{p}_{u=1} 
  \boldsymbol{\Phi}_{u}
  \boldsymbol{\eta}_{t-u} +
  \sum^{q}_{u=1} 
  \boldsymbol{\Theta}_{u}
  \boldsymbol{\zeta}_{t-u} +
   \boldsymbol{\Gamma}\mathrm{x}_{t} +
  \boldsymbol{\zeta}_\mathrm{t}
  \label{lv.eqn}
\end{align}

\noindent where (\ref{ov.eqn}) describes the observed variable or measurement model and (\ref{lv.eqn}) describes a vector autoregressive moving average (VARMA) latent time series. In the measurement model $\mathbf{y}_{t}$ is a $k\times1$ vector of observations at time $t$,  $\boldsymbol{\Lambda}_{u}$ is a sequences of $k \times m$ factor loading matrices up to order $s$, $\boldsymbol{\eta}_{t}$ is an $m\times1$ vector of latent factors at time $t$, and $\boldsymbol{\varepsilon}_{t}$ is a $k\times 1$ vector of unique factors at time $t$, and $\Cov(\boldsymbol{\varepsilon}_{t})=\boldsymbol{\Xi}$. For the latent variable time series $\boldsymbol{\eta}_{t}$,  $\boldsymbol{\Phi}_\mathrm{u}$ is a series of $m\times m$ matrices up to order $p$ containing the autoregressive and cross-regressive weights, $\boldsymbol{\Theta}_{u}$ is a series of $\emph{m}\times\emph{m}$ matrices up to order $q$ containing the moving average weights, $\boldsymbol{\Gamma}$ is a $\emph{m}\times\emph{r}$ matrix of regression coefficients relating an $r\times1$ vector of exogenous covariates, $\mathbf{x}_{t}$ to the latent series $\boldsymbol{\eta}_{t}$, and $\boldsymbol{\zeta}_{t}$ is a $m\times1$ vector of random shocks or innovations with $\Cov(\boldsymbol{\zeta}_{t})= \boldsymbol{\Psi}$.

\subsection{Nonlinear Dynamic Factor Model Specification}

To allow for arbitrarily time-varying parameters in \eqref{ov.eqn} and \eqref{lv.eqn} it is convenient to reformulate the linear dynamic factor model described above into the state space framework. Here we must also make the distinction between parameters which are time-invariant and those which we believe to vary in time.  Let the the column vector $\boldsymbol{\omega}$ contain the time-varying parameters from \eqref{ov.eqn} and \eqref{lv.eqn} such that $\boldsymbol{\omega}_{t}^{'} = [\upsilon(\boldsymbol{\Lambda})^{'}, \upsilon(\boldsymbol{\Phi})^{'},\upsilon(\boldsymbol{\Theta})^{'},\upsilon(\boldsymbol{\Gamma})^{'}]$. Here the $\upsilon(\cdot)$ operator stacks the unique time-varying (time-invariant) elements of each patterned matrix column-wise as in \citet{magnus1983}.  A number of options are available for modeling $\boldsymbol{\omega}_{t}$. Here we present the rather general specification
\begin{align}
\boldsymbol{\omega}_{t} &= 
  \boldsymbol{\omega}_{t-1} +
  \boldsymbol{\xi}_{t}
  \label{omega.eqn}
\end{align}
where $\boldsymbol{\xi}_{t}$ is a zero-mean white-noise process. This specification is equivalent to letting $\boldsymbol{\omega}_{t}$ obey a random walk. We note the specification of \eqref{omega.eqn} means $\boldsymbol{\omega}_{t}$ can vary arbitrarily across time without requiring  any pre-defined parametric representation.  This means $\boldsymbol{\omega}_{t}$ does not need to change linearly, quadratically, or obey any specific functional form. However, as the variance of  The only requirement placed on $\boldsymbol{\omega}_{t}$ is that it varies slowly in time relative to the variation observed in $\mathbf{y}_{t}$ \citep{priestley1988}. \textcolor{black}{One potential disadvantage related to this specification is the implicit assumption that the variance of $\boldsymbol{\omega}_{t}$ increases over time. However, the variance of  $\boldsymbol{\omega}_{t}$ may be of less interest when employing a time-varying parameter model, as is the case here. In practice there are also methods available to ensure the EKF yields consistent estimates of the time-varying states under this specification \citep[][p. 482]{bar2002}, which is often the primary objective of a time-varying parameter analysis.  Although other specifications are available, the specification in \ref{omega.eqn} allows for a parsimonious and easily interpretable characterization  of smoothly evolving parameters in the state-space framework. }

Now, let $\boldsymbol{\eta}^{*}_{t}$ represent an augmented state vector that has been expanded to include the original state variables in $\boldsymbol{\eta}_{t}$ as well as the time-varying parameters in $\boldsymbol{\omega}_{t}$, such that, 
\begin{align}
\boldsymbol{\eta}^{*}_{t} &= 
\begin{bmatrix}
    \boldsymbol{\eta}_{t}  \\
    \boldsymbol{\omega}_{t}
\end{bmatrix}
  \label{eta2.eqn}.
\end{align}
A nonlinear state space model for the augmented state vector in \eqref{eta2.eqn} and the observed variable time series can then be written as
\begin{align}
\mathbf{y}_{t} &= 
h(
  	\boldsymbol{\eta}^{*}_{t},
	\boldsymbol{\pi}
  ) +
  \tilde{\boldsymbol{\varepsilon}}_{t}
  \label{ssy.eqn}
  \\
\boldsymbol{\eta}^{*}_{t} &=
f(
  	\boldsymbol{\eta}^{*}_{t-1},
	\mathbf{x}_{t},
	\boldsymbol{\pi}
  ) +
  \tilde{\boldsymbol{\zeta}}_{t}
  \label{ssn.eqn}
\end{align}
where $\tilde{\boldsymbol{\varepsilon}}_{t}$ contains the adjusted elements from $\boldsymbol{\varepsilon}_{t}$, and similarly $\tilde{\boldsymbol{\zeta}}_{t}$ contains the adjusted elements in $\boldsymbol{\zeta}_{t}$ and $\boldsymbol{\xi}_{t}$. Similar to above the measurement and process noise vectors are normally distributed with mean zero and covariance matrices given by $\tilde{\boldsymbol{\Xi}}$ and $\tilde{\boldsymbol{\Psi}}$, respectively, and  $\boldsymbol{\pi}$ contains the time-invariant parameters from the following model matrices, 
\begin{align}
\boldsymbol{\pi}^{'} = [\upsilon(\boldsymbol{\Lambda})^{'}, \upsilon(\boldsymbol{\Phi})^{'},\upsilon(\boldsymbol{\Theta})^{'},\upsilon(\boldsymbol{\Gamma})^{'},\upsilon(\boldsymbol{\Xi})^{'}, \upsilon(\boldsymbol{\Psi})^{'}].
\end{align}
We assume the set of linear or nonlinear functions $h()$ and $f()$ describing the measurement relations and dynamic evolution of the augmented state vector to be continuously differentiable.

\section{The Extended Kalman Filter}

The Extended Kalman Filter \citep[EKF;][]{gelb2001,bar2002} is an extension of the classic Kalman Filter (KF) to the case of nonlinear dynamics and measurement processes. The EKF is appropriate for estimating the types of nonlinear state space models with additive noise described above. Unlike the traditional KF the EKF requires a series expansion of the nonlinearities in both the dynamics and measurement equations as a means to approximate the joint distribution of the latent states and observed variables. Here we employ the second-order EKF and thus include a second-order expansion to provide higher-order correction terms in the prediction and updating equations.

\subsection{Estimation Algorithm}

A single cycle of the KF algorithm can be understood as a mapping of the conditional mean and covariance of the states at time $t$ to the corresponding quantities at $t+1$ using the information set available at $t$.   Here the information set, $\mathcal{I}_{t}=(\mathbf{Y}_{t}, \mathbf{X}_{t-1})$, includes $\mathbf{Y}_{t}$, or the sequence of observations through time $t$ (including the initial state) and $\mathbf{X}_{t-1}$, the sequence of known exogenous inputs prior to time $t$. Using this prior information we can define the approximate conditional mean of the state as $\hat{\boldsymbol{\eta}}^{*}_{j|k}\approx \mathbb{E}(\boldsymbol{\eta}^{*}_{j}|\mathcal{I}_{k})$, the estimator error as $\tilde{\boldsymbol{\eta}}^{*}_{j|k}= \boldsymbol{\eta}^{*}_{j}-\hat{\boldsymbol{\eta}}^{*}_{j|k}$, and the associated conditional covariance matrix of the state (or in the case of a nonlinear model the covariance matrix of the estimation error) as $\mathbf{P}_{j|k}=\mathbb{E}[\tilde{\boldsymbol{\eta}}^{*}_{j|k}\tilde{\boldsymbol{\eta}}^{*'}_{j|k}|\mathcal{I}_{k}]$. Here we note that under differing conditions the conditional mean will represent an estimate of the state if $j=k$, an estimate of the smoothed state if $j < k$, and an estimate of the predicted state if $j > k$.
\subsection{Initial State and Design Parameters}
The EKF algorithm requires a number of quantities prior to its initialization.  These quantities include an initial estimate of the state, $\boldsymbol{\eta}^{*}_{0|0}$, and the corresponding state covariance matrix, $\mathbf{P}_{0|0}$.  In addition, estimates of the measurement noise covariance matrix, $ \tilde{\boldsymbol{\Xi}}_{t}$, and the process noise covariance, $ \tilde{\boldsymbol{\Psi}}_{t}$, are required. The choice of these quantities is far from trivial and can have a large impact on the subsequent performance of the filter. In previous research on the EKF these design parameters have often been chosen arbitrarily making it difficult to assess the performance of the estimator against alternative algorithms \citep[see][]{schneide2013}. For this reason we discuss the choice of these parameters in the context of the current problem. First, sufficiently precise estimates of both $\boldsymbol{\eta}^{*}_{0|0}$ and $\mathbf{P}_{0|0}$ (as well as $\boldsymbol{\pi}$) can be obtained from a preliminary P-Technique factor analysis \citep{molenaar2009}. For the elements of $\mathbf{P}_{0|0}$ pertaining to the time-varying parameters \citet[p. 482]{bar2002} suggest using a few percent of the estimated time-invariant coefficient pertaining to the suspected time-varying parameter as an initial estimate of the process noise variance.

The choice of $\tilde{\boldsymbol{\Xi}}$ and $ \tilde{\boldsymbol{\Psi}}$ is considerably more difficult.  In the context of the SEKF with time-invariant parameters a number of methods have been suggested. These approaches could be adapted to the context of time-varying parameters, however, the utility of this approach has not been demonstrated in the literature.  Here we adopt a procedure proposed by \citep{molenaar2016} for tuning $\tilde{\boldsymbol{\Xi}}$ and $ \tilde{\boldsymbol{\Psi}}$ along with the other time-invariant parameters in $\boldsymbol{\pi}$ using the raw data log-likelihood function
\begin{align}
\mathrm{log}\: L(\boldsymbol{\pi}) &= \frac{1}{2}
\sum^{T}_{t=1}-k\:\mathrm{log}(2\pi)-\mathrm{log}(|\mathbf{S}_{t}|) + \tilde{\mathbf{y}}_{t}^{'}\mathbf{S}_{t}^{-1}\tilde{\mathbf{y}}_{t}
\label{logl}
\end{align}
where $\tilde{\mathbf{y}}_{t}$ contains the one-step ahead prediction errors obtained from the SEKF and $\mathbf{S}_{t}$ is the corresponding covariance matrix. Here we obtain parameter estimates based on the assumptions of a linear measurement model with additive and Gaussian distributed process and measurement noises. If the one-step ahead prediction errors are normally and independently distributed after removing the time dynamics implied by the model the optimization procedure will yield maximum likelihood estimates \citep{chow2007c}. In addition, the tuning procedure described above has the added benefit that the estimated values of $ \tilde{\boldsymbol{\Psi}}$ returned by this index the variability of the time-varying parameters themselves, allowing for the possibility the parameters are in fact time-invariant (e.g. zero variance). Once this log-likelihood function has been optimized with respect to all the time invariant parameters in $\boldsymbol{\pi}$ these estimates are treated as fixed to obtain smoothed state estimates.  This procedure is described in greater detail below.

\subsection{Square Root Filter}

A number of authors have analyzed square root versions of the first-order EKF \citep{park1995, chandra2019}.  However, to the best of our knowledge we are the first to explicitly detail how a square root filter can be adapted to the second-order EKF. In the square-root filter described here the square root of the state covariance matrix, rather than the state covariance matrix itself, is propagated through the Kalman recursions.  Generally, the structure of the square root covariance matrix allows for a number of improvements over the standard implementation, including (a) assurance of a symmetric positive definite error covariance matrix, (b) higher order precision and therefore improved numerical accuracy \citep{grewal2001}, and (c) improved performance in parallel implementations \citep{park1995}.

\subsection{State Prediction}

As stated previously the primary objective of the SEKF is to obtain unbiased estimates of the state vectors, or latent variables, $\boldsymbol{\eta}^{*}_{t}$, by minimizing the least squares prediction error.  At each time period the EKF completes two steps: (1) In the \emph{prediction} step, a model-based prediction of the individual scores $\hat{\boldsymbol{\eta}}^{*}_{t|t-1}$ is obtained from scores at the previous time point $\boldsymbol{\eta}^{*}_{t-1|t-1}$. (2) In the \emph{correction} step, the model-based prediction is corrected using observed information gathered at time $t$. These two alternating steps occur in an on-line fashion at each time point, moving through the data structure sequentially. 

To obtain the predicted state $\hat{\boldsymbol{\eta}}^{*}_{t|t-1}$ we expand in Taylor Series the nonlinear function given in \eqref{ssn.eqn} around the previous state estimate $\hat{\boldsymbol{\eta}}^{*}_{t-1|t-1}$. To the second order this series expansion ignoring the higher order terms is given by
\begin{IEEEeqnarray}{rCl}
\boldsymbol{\eta}^{*}_{t|t-1} &=&
f(\boldsymbol{\eta}^{*}_{t-1|t-1},\mathbf{x}_{t-1},\boldsymbol{\pi}) + 
 \frac{\partial f(
 	\boldsymbol{\eta}^{*}_{t-1|t-1},
\mathbf{x}_{t-1},
\boldsymbol{\pi}
 )}{\partial\: 
 	\boldsymbol{\eta}^{*}_{t}
 }[\boldsymbol{\eta}^{*}_{t-1|t-1} - \hat{\boldsymbol{\eta}}^{*}_{t-1|t-1}] \: + \nonumber \\
&& \frac{1}{2} \sum_{i=1}^{m^{*}} \mathbf{e}_{i}[\boldsymbol{\eta}^{*}_{t-1|t-1} - \hat{\boldsymbol{\eta}}^{*}_{t-1|t-1}]^{'}
 \frac{\partial^{2} f_{i}(
 	\boldsymbol{\eta}^{*}_{t-1|t-1},
\mathbf{x}_{t-1},
\boldsymbol{\pi}
 )}{\partial\: 
 	\boldsymbol{\eta}^{*2}_{t}
 }[\boldsymbol{\eta}^{*}_{t-1|t-1} - \hat{\boldsymbol{\eta}}^{*}_{t-1|t-1}] + \nonumber \\
 && +
  \tilde{\boldsymbol{\zeta}}_{t}.
 \label{statepred0}
\end{IEEEeqnarray}
where $\partial f(\boldsymbol{\eta}^{*}_{t-1|t-1}, \mathbf{x}_{t-1},\boldsymbol{\pi} ) / \partial\boldsymbol{\eta}^{*}_{t}$ is the Jacobian of the vector function $f$ evaluated at the previous estimate of the state, and $\partial^2 f_{i}(\boldsymbol{\eta}^{*}_{t-1|t-1}, \mathbf{x}_{t-1},\boldsymbol{\pi} ) / \partial \boldsymbol{\eta}^{*2}_{t}$ is the Hessian matrix of the $i^{th}$ state variable. Moving forward we will use $\mathbf{J}_{f}$ and $\mathbf{H}_{f,i}$ to represent the Jacobian and Hessian of the state, respectively. Furthermore, we take the values of $\boldsymbol{\pi}$ to be known within the recursions of the EKF. 

Taking the conditional expectation of  \eqref{statepred0} given the observed information and known parameter values we obtain the following formula for the predicted state
\begin{align}
\hat{\boldsymbol{\eta}^{*}}_{t|t-1} &=
f(
 	\boldsymbol{\eta}^{*}_{t-1|t-1},
\mathbf{x}_{t-1},
\boldsymbol{\pi}
 ) + 
 \frac{1}{2}
 \sum^{m^{*}}_{i=1}
 \mathbf{e}_{i}
 \mathrm{tr}\big[
  \bar{\mathbf{P}}_{t-1|t-1}^{'}
\mathbf{H}_{f,i}
 \bar{\mathbf{P}}_{t-1|t-1}\big],
 \label{statepred1}
\end{align}
as only $\boldsymbol{\eta}^{*}$ and $\tilde{\boldsymbol{\zeta}}$ are random variables in \eqref{statepred0} and the second order term is simplified based on the expectation of quadratic forms.  The positive semi-definite state covariance matrix $\mathbf{P}$ has been factorized such that $\bar{\mathbf{P}} = \mathbf{P}^{1/2}$, $\bar{\mathbf{P}}^{'} = \mathbf{P}^{'1/2}$, and  $\mathbf{P}= \mathbf{P}^{1/2} \mathbf{P}^{'1/2}$. \textcolor{black}{For the positive semi-definite matrix $\mathbf{P}$ the Cholesky decomposition is a computationally convenient method that satisfies the identity above.} Furthermore, by subtracting \eqref{statepred1} from \eqref{statepred0} we obtain the state prediction error $\tilde{\boldsymbol{\eta}}^{*}$ as 
\begin{IEEEeqnarray}{rCl}
\tilde{\boldsymbol{\eta}}^{*}_{t|t-1} &=&
\mathbf{J}_{f}
 \tilde{\boldsymbol{\eta}}^{*}_{t-1|t-1} +
  \frac{1}{2} \sum_{i=1}^{m^{*}} \mathbf{e}_{i}
  \Big[
   \tilde{\boldsymbol{\eta}}^{*}_{t-1|t-1} 
   \mathbf{H}_{f,i}
  \tilde{\boldsymbol{\eta}}^{*'}_{t-1|t-1} -
  \mathrm{tr}(   \bar{\mathbf{P}}_{t-1|t-1}^{'}
\mathbf{H}_{f,i}
 \bar{\mathbf{P}}_{t-1|t-1})
  \Big] +  \tilde{\boldsymbol{\zeta}}_{t},
 \label{errpred}
\end{IEEEeqnarray}
\textcolor{black}{where again the higher order terms are ignored and $\mathbf{e}_{i}$ is a Cartesian basis vector with the $i$th element set to unity and the remaining elements set to zero.} \\
\subsection{State Prediction Covariance Matrix}
In the standard implementation of the SEKF the mean squared error covariance matrix of \eqref{errpred} is written as
\begin{IEEEeqnarray}{rCl}
\mathbf{P}_{t|t-1} &=&
\mathbf{J}_{f}\mathbf{P}_{t-1|t-1}\mathbf{J}_{f}^{'} +
\frac{1}{2} \sum_{i=1}^{m^{*}}  \sum_{j=1}^{m^{*}}\mathbf{e}_{i}\mathbf{e}_{j}^{'}
\mathrm{tr}\big[
\mathbf{H}_{f,i}\mathbf{P}_{t-1|t-1}\mathbf{H}_{f,j}^{'}\mathbf{P}_{t-1|t-1}
\big] + \tilde{\boldsymbol{\Psi}}_{t-1}
 \label{statecov}
\end{IEEEeqnarray}
where use has been made of results for the covariance matrix of quadratic forms and the assumption that moments greater than two are equal to zero. However, in the square root version of the filter we are are only interested in updating $\bar{\mathbf{P}}$. In the square root version of a first-order EKF this is straightforward as only the factorized version of $\tilde{\boldsymbol{\Psi}}_{t-1}$ is required. However, in the SEKF we must also include the linearization error in our factorization such that
\begin{IEEEeqnarray}{rCl}
\mathbf{P}_{\eta_\varepsilon} &=&
\frac{1}{2} \sum_{i=1}^{m^{*}}  \sum_{j=1}^{m^{*}}\mathbf{e}_{i}\mathbf{e}_{j}^{'}
\mathrm{tr}\big[
\bar{\mathbf{P}}_{t-1|t-1}^{'}\mathbf{H}_{f,i}\bar{\mathbf{P}}_{t-1|t-1}
\bar{\mathbf{P}}_{t-1|t-1}^{'}\mathbf{H}_{f,i}\bar{\mathbf{P}}_{t-1|t-1}
\big] + \tilde{\boldsymbol{\Psi}}_{t-1},
 \label{linerrmat2}
\end{IEEEeqnarray}
where here and below a matrix $\mathbf{P}_{\varepsilon} $ is factorized using the Cholesky decomoposition such that $\bar{\mathbf{P}}_{\varepsilon} = \mathbf{P}_{\varepsilon}^{1/2}$, $\bar{\mathbf{P}}_{\varepsilon}^{'} = \mathbf{P}_{\varepsilon}^{'1/2}$, and  $\mathbf{P}_{\varepsilon}= \mathbf{P}_{\varepsilon}^{1/2} \mathbf{P}_{\varepsilon}^{'1/2}$. With these quantities in hand the factorized state prediction covariance matrix can be obtained from the following identity and transformation
\begin{IEEEeqnarray}{rCl}
\boldsymbol{\kappa}
 \big[
\mathbf{J}_{f}\bar{\mathbf{P}}_{t-1|t-1}
\quad 
\bar{\mathbf{P}}_{\eta_\varepsilon}
\big]^{'} 
 &=& 
\bar{\mathbf{P}}_{t|t-1}^{'} 
 \label{factorizedstatpred}
\end{IEEEeqnarray}
where $\boldsymbol{\kappa}$ indicates the transformation of a symmetric matrix to upper-triangular via the QR decomposition. 
\subsection{Measurement Prediction}
We can write the predicted measurement observations $\hat{\mathbf{y}}_{t|t-1}$ for the second-order EKF as 
\begin{IEEEeqnarray}{rCl}
\hat{\mathbf{y}}_{t|t-1} &=& h(\boldsymbol{\eta}^{*}_{t|t-1},\boldsymbol{\pi}) +
 \frac{1}{2}\sum^{k}_{i=1}
 \mathbf{e}_{i}
 \mathrm{tr}\big[
 \bar{\mathbf{P}}_{t|t-1}
\mathbf{H}_{h,i}
\bar{\mathbf{P}}^{'}_{t|t-1}\big]
 \label{measpred}
\end{IEEEeqnarray}
using the factorized state prediction covariance matrix where the associated measurement prediction covariance matrix is given by 
\begin{IEEEeqnarray}{rCl}
\mathbf{S}_{t|t-1} &=& 
\mathbf{J}_{h}\bar{\mathbf{P}}^{'}_{t|t-1}\bar{\mathbf{P}}_{t|t-1}\mathbf{J}_{h}^{'} +  
\frac{1}{2} \sum_{i=1}^{k}  \sum_{j=1}^{k}\mathbf{e}_{i}\mathbf{e}_{j}^{'}
\mathrm{tr}\big[\bar{\mathbf{P}}^{'}_{t|t-1}
\mathbf{H}_{h,i}\bar{\mathbf{P}}_{t|t-1}\bar{\mathbf{P}}^{'}_{t|t-1}\mathbf{H}_{h,j}^{'}\bar{\mathbf{P}}_{t|t-1}
\big] + \tilde{\boldsymbol{\Theta}}_{t-1} \nonumber
 \label{meascov}
\end{IEEEeqnarray}
and $\mathbf{J}_{h}=\partial h(\boldsymbol{\eta}^{*}_{t|t-1},\boldsymbol{\pi} ) / \partial \boldsymbol{\eta}^{*}_{t}$ is the Jacobian of the vector function $h$ evaluated at the updated estimate of the state, and $\mathbf{H}_{h,i} = \partial^2 h_{i}(\boldsymbol{\eta}^{*}_{t|t-1}, \boldsymbol{\pi} ) / \partial \boldsymbol{\eta}^{*2}_{t}$ is the Hessian matrix of the $i^{th}$ state variable.

\subsection{State Update Equations}
Traditionally in the non-square-root second-order filter we compute the state update, $\hat{\boldsymbol{\eta}}^{*}_{t|t}\approx \mathbb{E}(\boldsymbol{\eta}^{*}_{t}|\mathcal{I}_{t-1})$, directly from 
\begin{IEEEeqnarray}{rCl}
\label{stateupdate}
\hat{\boldsymbol{\eta}}^{*}_{t|t} & = & \hat{\boldsymbol{\eta}}^{*}_{t|t-1} + \mathbf{W}_{t}\tilde{\mathbf{y}}_{t}
\end{IEEEeqnarray}
where $\mathbf{W}_{t}  = \mathbf{P}_{t|t-1}\mathbf{J}_{h}^{'}\mathbf{S}_{t}^{-1}$ is the filter gain and $\tilde{\mathbf{y}}_{t}=\mathbf{y}_{t}-\hat{\mathbf{y}}_{t}$ is the measurement residual, and the updated state covariance is $
\mathbf{P}_{t|t} = \mathbf{P}_{t|t-1} - \mathbf{W}_{t}\mathbf{S}_{t}\mathbf{W}_{t}^{'}$. In the modified square-root version of the filter proposed here we use the factorized state prediction covariance matrix to construct the following identity
\begin{IEEEeqnarray}{rCl}
\label{stateupdate2}
\boldsymbol{\kappa}
\begin{bmatrix}
 \bar{\tilde{\boldsymbol{\Theta}}}^{'}_{t-1} & \mathbf{0} \\
 \bar{\mathbf{P}}_{t|t-1}\mathbf{J}_{h}^{'} & \bar{\mathbf{P}}^{'}_{t|t-1}
 \end{bmatrix}
 & = &
 \begin{bmatrix}
 \boldsymbol{\Delta}_{t} & \boldsymbol{\Upsilon}_{t} \\
\mathbf{0} &  \bar{\mathbf{P}}_{t|t}
 \end{bmatrix}
\end{IEEEeqnarray}
where  $\bar{\tilde{\boldsymbol{\Theta}}}^{'}$ is the square root of the measurement noise covariance matrix, $\boldsymbol{\kappa}$ again indicates the transformation of a symmetric matrix to upper-triangular via the QR decomposition and the filter gain is calculated by $\mathbf{W}_{t} = \boldsymbol{\Upsilon}_{t} \boldsymbol{\Delta}_{t}^{-1}$.  The updated state can then be obtained using \eqref{stateupdate}.\\
Thus far we have described a single cycle of the SR-SEKF algorithm.  This cycle is then repeated for each subsequent set of observations from $t=2$ to $T$, until the state vector $\hat{\boldsymbol{\eta}^{*}}_{t|t}$ has been estimated for each $t$ in the series.  This full set of recursions produces a likelihood value based on the prediction errors as summarized in Equation \ref{logl}, which is then optimized with respect to the time-invariant parameters in the model. Included in this set of time-invariant parameters being optimized are the diagonal elements of $ \tilde{\boldsymbol{\Psi}}$, which in the case of the time-varying parameters provide additional information as to whether or not the parameter varies meaningfully across time \citep{molenaar2016}. Finally, once the likelihood function has been optimized the estimated time-invariant parameters are treated as fixed and the states can be smoothed to obtain improved estimates of the states and state covariance matrix. \textcolor{black}{Although a detailed analysis of the computational complexity of our EKF implementation is beyond the scope of this paper interested readers can find a detailed discussion of computational complexity reduction methods for filtering algorithms in \citet{raitoharju2019}.}
\subsection{Fixed-Interval Smoother}
Smoothing is the estimation of the state at time $j$ with an interval of data where $j < k$. A number of smoothers have been defined based on the selection of $k$, however, we only consider the fixed-interval smoother where $k=T$. Fixed-interval smoothing is equivalent to smoothing the entire trajectory of the estimated states based on all of the available timepoints. Here we implement the Rauch-Tung-Striebel smoother \citep[RTS;][]{rauch1965} adapted to the SEKF. This fixed-interval smoother requires the following quantities to be saved from the forward recursions of the SEKF: $\hat{\boldsymbol{\eta}}^{*}_{t|t}$, $\hat{\boldsymbol{\eta}}^{*}_{t|t-1}$, $\hat{\mathbf{P}}_{t|t}$, and $\hat{\mathbf{P}}_{t|t-1}$.  Using these estimates calculated backwards in time starting from $k = T$, the smoothed state estimates are calculated as
\begin{IEEEeqnarray}{rCl}
\hat{\boldsymbol{\eta}}^{*}_{t|N} & = & \hat{\boldsymbol{\eta}}^{*}_{t|t} + 
\mathbf{C}_{t}[
 \hat{\boldsymbol{\eta}}^{*}_{t+1|N} - \hat{\boldsymbol{\eta}}^{*}_{t+1|t} ]
\label{smoothedstate}
\end{IEEEeqnarray}
where
\begin{IEEEeqnarray}{rCl}
\mathbf{C}_{t} & = & 
\mathbf{P}_{t}\mathbf{J}_{f}^{'}\mathbf{P}_{t|t-1}^{-1}
\label{smoothedc}
\end{IEEEeqnarray}
and the smoothed state covariance matrix is
\begin{IEEEeqnarray}{rCl}
\mathbf{P}_{t|N} & = & 
\mathbf{P}_{t} + \mathbf{C}_{t} [\mathbf{P}_{t+1|N}-\mathbf{P}_{t|t-1}]\mathbf{C}_{t}^{'}.
\label{smoothedc}
\end{IEEEeqnarray}

\section{Monte Carlo Simulations}

Monte Carlo simulations were conducted to examine the finite-sample properties of the SEKF estimator in nonlinear time-varying parameter state space models. The simulations were loosely based on the empirical example considered in this paper, where daily measures of two latent constructs (positive and negative affect) were collected prior to, during, and following a multiple-week intervention. A number of scenarios regarding the time-varying dynamics and hypothesized intervention effects could be considered. For the purpose of this paper we focused our simulations on two situations we thought were substantively interesting while also having the potential to contribute new information to the literature on time-varying parameter estimation.  First, we considered the scenario where the cross-regressive effects between the latent states (positive and negative affect) varied across time.  Second, we investigated the case where the effect of an intervention on the latent states (positive and negative affect) varied in time following the onset of the intervention, possibly continuing even after the intervention ceased. These two scenarios are represented in simulations 1 and 2, respectively. 

As previously mentioned, in addition to the overall performance of the estimator in terms of parameter recovery we were also interested in answering a number of preliminary questions relevant to the use of these models in psychological research.  First, is the estimator able to accurately recover multiple time-varying parameters when they exist in the data generating model?  Second, what is the impact of allowing many model parameters to vary in the estimated model when only a subset of those parameters are time-varying in the true or data generating model?  

\textcolor{black}{Finally, we also wanted to better understand the potential advantages of our proposed square-root algorithm (SR-SEKF) over the standard second-order EKF (SEKF). In theory these two approaches are algebraically equivalent, however, previous research has shown square-root filters can reduce the computational complexity compared to standard algorithms, with downstream effects on computational efficiency, convergence rates and parameter estimation.  For these reasons we also chose to compare our approach to its non-square-root counterpart.} 

All the estimation routines described in this paper and employed for the simulations and empirical example were coded in R by the first author. In the remainder of this section we describe in more detail the outcomes of interest, data generating models, and simulation results.  

\subsection{Measures}
To compare the performance of the estimator across time-varying and time-invariant parameters we examined its relative bias and efficiency within each simulation condition. For parameters that were generated as time-invariant and also treated as time-invariant in the fitted model the mean relative bias was calculated as $[\sum^{N}_{k=1} (\hat{\pi}_{ak}-\pi_{a})/\pi_{a}]/N$ where $\pi_{a}$ is the data generating parameter in a given simulation condition, $\hat{\pi}_{ak}$ is the estimate for parameter $a$ in the $k^{th}$ Monte Carlo replication, and $N$ is the total number of replications. \textcolor{black}{For parameters that were generated as time-invariant but treated as time-varying in the fitted model mean relative bias was calculated as $\sum^{T}_{t=1}\left([\sum^{N}_{k=1} (\hat{\pi}_{akt}-\pi_{a})/\pi_{a}]/N\right)/T$ where $\hat{\pi}_{akt}$ is the estimate for parameter $a$ at timepoint $t$ in the $k^{th}$ Monte Carlo replication.} Mean relative bias was then multiplied by 100 to obtain the mean percentage of relative bias. Second, for parameters generated and estimated as time-invariant we examined the variability associated with each model parameter using the standard deviation of $\hat\pi_{a}$ across replications within a single block of the simulation, $SD(\hat\pi_{ak})$.  We also report the accuracy of the estimation procedure for detecting whether a parameter was correctly time-varying or time-invariant based on the conservative measure of zero variance for the final smoothed estimates of each parameter within a given sample. 
 
\subsection{Data Generating Models}

As stated previously we considered two main simulation conditions. First, we examined the case where the cross-regressive effects between the latent states (positive and negative affect) vary across time.  Second, we investigated the case where the effect of an intervention on the latent states (positive and negative affect) varied in time.   

\begin{table}[ht]
\centering
\caption{Taxonomy of Simulation Conditions and Parameter Designation} 
\label{tab:sim}
\scalebox{.7}{
\begin{tabular}{ccllll}
  \toprule
 &&    \multicolumn{2}{c}{Data-Generating Model} &
  \multicolumn{2}{c}{Fitted Model}  \\
   \cmidrule(lr){3-4} 
   \cmidrule(lr){5-6} 
  &&    \multicolumn{2}{c}{Parameter Type} &
  \multicolumn{2}{c}{Parameter Type}  \\
     \cmidrule(lr){3-4} 
   \cmidrule(lr){5-6} 
\multicolumn{1}{c}{Simulation} &
\multicolumn{1}{c}{Condition} &
\multicolumn{1}{c}{Time-Varying} &
  \multicolumn{1}{c}{Time-Invariant} &
   \multicolumn{1}{c}{Time-Varying} &
  \multicolumn{1}{c}{Time-Invariant} \\
  \midrule
  1 & A & $\Phi_{12}$, $\Phi_{21}$ 
  	& $\Phi_{11}$, $\Phi_{22}$, $\Gamma_{1}$, $\Gamma_{2}$, 
	$\boldsymbol{\Lambda}$,
	$\tilde{\boldsymbol{\Xi}}$, $\tilde{\boldsymbol{\Psi}}$	
  	& $\Phi_{12}$, $\Phi_{21}$ 
	&  $\Phi_{11}$, $\Phi_{22}$, $\Gamma_{1}$, $\Gamma_{2}$,
	$\boldsymbol{\Lambda}$,   
	$\tilde{\boldsymbol{\Xi}}$, $\tilde{\boldsymbol{\Psi}}$\\
   1 & B & $\Phi_{12}$, $\Phi_{21}$ 
  	& $\Phi_{11}$, $\Phi_{22}$, $\Gamma_{1}$, $\Gamma_{2}$, 
	$\boldsymbol{\Lambda}$,
	$\tilde{\boldsymbol{\Xi}}$, $\tilde{\boldsymbol{\Psi}}$	    &$\Phi_{12}$, $\Phi_{21}$, $\Phi_{11}$, $\Phi_{22}$ &$\Gamma_{1}$, $\Gamma_{2}$,
	$\boldsymbol{\Lambda}$,   
	$\tilde{\boldsymbol{\Xi}}$, $\tilde{\boldsymbol{\Psi}}$ \\
   1 & C & $\Phi_{12}$, $\Phi_{21}$ 
  	& $\Phi_{11}$, $\Phi_{22}$, $\Gamma_{1}$, $\Gamma_{2}$, 
	$\boldsymbol{\Lambda}$,
	$\tilde{\boldsymbol{\Xi}}$, $\tilde{\boldsymbol{\Psi}}$	    & $\Phi_{12}$, $\Phi_{21}$, $\Phi_{11}$, $\Phi_{22}$, $\Gamma_{1}$, $\Gamma_{2}$&
	$\boldsymbol{\Lambda}$,   
	$\tilde{\boldsymbol{\Xi}}$, $\tilde{\boldsymbol{\Psi}}$ \\
    2 & A & $\Gamma_{1}$, $\Gamma_{2}$ & $\Phi_{12}$, $\Phi_{21}$, $\Phi_{11}$, $\Phi_{22}$, 
	$\boldsymbol{\Lambda}$,
	$\tilde{\boldsymbol{\Xi}}$, $\tilde{\boldsymbol{\Psi}}$	 & $\Gamma_{1}$, $\Gamma_{2}$&  $\Phi_{12}$, $\Phi_{21}$, $\Phi_{11}$, $\Phi_{22}$,
	$\boldsymbol{\Lambda}$,   
	$\tilde{\boldsymbol{\Xi}}$, $\tilde{\boldsymbol{\Psi}}$\\
  2 & B & $\Gamma_{1}$, $\Gamma_{2}$ &$\Phi_{12}$, $\Phi_{21}$, $\Phi_{11}$, $\Phi_{22}$, 
	$\boldsymbol{\Lambda}$,
	$\tilde{\boldsymbol{\Xi}}$, $\tilde{\boldsymbol{\Psi}}$&$\Gamma_{1}$, $\Gamma_{2}$, $\Phi_{12}$, $\Phi_{21}$ &  $\Phi_{11}$, $\Phi_{22}$,
	$\boldsymbol{\Lambda}$,   
	$\tilde{\boldsymbol{\Xi}}$, $\tilde{\boldsymbol{\Psi}}$\\
  2 & C & $\Gamma_{1}$, $\Gamma_{2}$ &$\Phi_{12}$, $\Phi_{21}$, $\Phi_{11}$, $\Phi_{22}$, 
	$\boldsymbol{\Lambda}$,
	$\tilde{\boldsymbol{\Xi}}$, $\tilde{\boldsymbol{\Psi}}$  & $\Gamma_{1}$, $\Gamma_{2}$, $\Phi_{12}$, $\Phi_{21}$, $\Phi_{11}$, $\Phi_{22}$&
	$\boldsymbol{\Lambda}$,   
	$\tilde{\boldsymbol{\Xi}}$, $\tilde{\boldsymbol{\Psi}}$ \\
  \bottomrule
  \end{tabular}
}
\end{table}

Continuous multivariate latent time series were generated in accordance with equations \eqref{ov.eqn} and \eqref{lv.eqn} for time series lengths of $T=70,200,500$. Random shock vectors and measurement errors were generated from $\mathcal{N}(\bm{0},\bm{\Psi})$ and $\mathcal{N}(\bm{0},{\boldsymbol{\Xi} })$, respectively. A burn-in period of 1,000 time points was discarded from the series to attenuate any persisting effects of the initial parameters.  100 datasets were generated for each block of the simulation design. Across our two main simulation conditions the data was generated with factor loadings equal to $(\Lambda_{11},\Lambda_{21},\Lambda_{31},\Lambda_{42},\Lambda_{52},\Lambda_{62} ) = (1,1,1,1,1,1)$, measurement error variances equal to $(\Xi_{11},\Xi_{22},\Xi_{33},\Xi_{44},\Xi_{54},\Xi_{66} ) = (.2,.2,.2,.2,.2,.2)$, and autoregressive coefficients as $(\Phi_{11},\Phi_{22}) = (.7,.5)$. Furthermore, the exogenous input variable $\mathbf{x}_{1}$ was coded as zero or one depending on whether the intervention was active at a given timepoint.  To mirror the empirical example, treatment began after approximately $2/11$ of the series had elapsed, and continued until the final observation point.  As the treatment in our empirical example involved teaching the subjects meditation techniques that were carried over to the home environment this setup was designed to reflect the uncertainty in the duration of treatment effects. Additional details regarding the individual model specifications used to generate and fit individual datasets will be described in detail below. 

In simulation 1 we sought to investigate whether the proposed estimator could accurately recover time-varying cross-regressive parameters $\Phi_{12}$ and $\Phi_{21}$. In the data generating model $\Phi_{12}$ and $\Phi_{21}$ were generated as random walks and then scaled to the range of $(-0.3,0.3)$. To better reflect what is likely a smooth evolution of the parameters in terms of daily changes in affect regulation we locally regressed and smoothed the generated random walks using a $50\%$ smoothing span \citep{cleveland1979}.  Visual depiction of the data generating values for the cross-regressive coefficients are shown as the "True States" in Figure~\ref{fig:sim1}. As the focus of simulation 1 was on the cross-regressive coefficients, the parameters describing the treatment effect on the latent constructs were held constant at $\Gamma_{1}=0.5$ and $\Gamma_{2}=0.5$.  As it will often the be the case that researchers do not know \emph{a priori} which parameters are time-varying we also wanted to better understand the consequences of allowing many parameters to vary in time.  For simulations 1 and 2 (presented below) this was examined across three sub-conditions  A, B and C. In sub-condition A, only $\Phi_{12}$ and $\Phi_{21}$ were estimated as time-varying and the remaining parameters were estimated as time-invariant.  In sub-condition B $\Phi_{12}$ and $\Phi_{21}$, as well as the autoregressive coefficients, $\Phi_{11}$ and $\Phi_{22}$, were estimated as time-varying. In sub-condition C, all the elements of $\boldsymbol{\Phi}$, as well as the treatment effects in $\boldsymbol{\Gamma}$ were estimated as possibly varying across time. 

In simulation 2 we examined whether the estimator could accurately recover time-varying treatment effects characterized by the $\Gamma_{1}$ and $\Gamma_{2}$ parameters. Just as in simulation 1 the time-varying parameters were generated as random walks, scaled to the range of $(-0.5,0.5)$ and smoothed. Visual depiction of the data generating values for the time-varying treatment effects are shown under the label "True States" in Figure~\ref{fig:sim2}.  In simulation 2 the cross-regressive parameters were held constant at $\Phi_{12}=-0.2$ and $\Phi_{21}=-0.3$. We also extended simulation 2 across the three sub-conditions.  In sub-condition A  $\Gamma_{1}$ and $\Gamma_{2}$ were estimated as time-varying and the remaining parameters were estimated as time-invariant. In sub-condition B, the treatment effects ($\Gamma_{1}, \Gamma_{2}$) and cross-regressive parameters ($\Phi_{12}, \Phi_{21}$) were estimated as time-varying, and in sub-condition C all the elements of $\boldsymbol{\Phi}$ and $\boldsymbol{\Gamma}$ were estimated as time-varying.  A taxonomy of the parameter types and their status across conditions and model designations (data-generating vs. estimated) is provided in Table~\ref{tab:sim}.

\subsection{Results}

\subsubsection{Relative Bias for Parameters Generated as Time-Invariant}

The percentage of relative bias for the parameters generated as time-invariant across all simulation conditions are given in Table~\ref{tab:bias}. Generally the bias in parameter estimates was small and rarely exceeded $5\%$, which is often used as a threshold of meaningful parameter bias. For the factor loadings and measurement error variances the bias never exceeded $4\%$, even in the smallest time series length of $T=70$. In simulation 1, where the cross-regressive parameters ($\Phi_{12}, \Phi_{21}$) were generated as time-varying, bias only exceeded $5\%$ for the exogenous treatment effects $\Gamma_{1}$ and $\Gamma_{2}$ and even here the bias was relatively small at $-8\%$ and $-6\%$, respectively. 

In simulation 2 where the treatment effect parameters, $\Gamma_{1}$ and $\Gamma_{2}$, were generated as time-varying, estimation of the smaller of the cross-regressive parameters, $\Phi_{12}$, and the autoregressive parameter $\Phi_{11}$ at the shortest time series length showed considerable bias compared to other parameters in the model.  In regards to our original question of whether allowing additional parameters to vary across time in the estimated model impacted the parameter recovery, we found no evidence of deleterious effects in terms of parameter bias. Bias in the the parameters generated and estimated as time-invariant did not increase as a function of increasing numbers of time-varying parameters.  Similarly, even parameters that were generated as time-invariant but estimated as time-varying showed little to no additional bias compared to the case when they were correctly estimated as time-invariant.

\begin{table}[ht]
\centering
\caption{Mean Percentage of Relative Bias for Parameters Generated as Time-Invariant} 
\label{tab:bias}
\scalebox{.7}{
\begin{threeparttable}
\begin{tabular}{rrrrrrrrrrrrrrrrrrr}
  \toprule
  &   \multicolumn{18}{c}{Time Series Length}  \\
 \cmidrule(lr){2-19} & \multicolumn{6}{c}{T = 70} &
  \multicolumn{6}{c}{T = 200} &
  \multicolumn{6}{c}{T = 500} \\
 \cmidrule(lr){2-7} \cmidrule(lr){8-13}\cmidrule(lr){14-19} & \multicolumn{6}{c}{Simulation Condition} &
  \multicolumn{6}{c}{Simulation Condition} &
  \multicolumn{6}{c}{Simulation Condition} \\
 \cmidrule(lr){2-7} \cmidrule(lr){8-13}\cmidrule(lr){14-19} &  \multicolumn{3}{c}{1} &
  \multicolumn{3}{c}{2} &
  \multicolumn{3}{c}{1} &
  \multicolumn{3}{c}{2} &
  \multicolumn{3}{c}{1} &
  \multicolumn{3}{c}{2} \\
 \cmidrule(lr){2-4} \cmidrule(lr){5-7}\cmidrule(lr){8-10}
  \cmidrule(lr){11-13} \cmidrule(lr){14-16}\cmidrule(lr){17-19} \multicolumn{1}{c}{Parameter} & 
  \multicolumn{1}{c}{A} &
  \multicolumn{1}{c}{B} &
  \multicolumn{1}{c}{C} &
  \multicolumn{1}{c}{A} &
  \multicolumn{1}{c}{B} &
  \multicolumn{1}{c}{C} &
  \multicolumn{1}{c}{A} &
  \multicolumn{1}{c}{B} &
  \multicolumn{1}{c}{C} &
  \multicolumn{1}{c}{A} &
  \multicolumn{1}{c}{B} &
  \multicolumn{1}{c}{C} &
  \multicolumn{1}{c}{A} &
  \multicolumn{1}{c}{B} &
  \multicolumn{1}{c}{C} &
  \multicolumn{1}{c}{A} &
  \multicolumn{1}{c}{B} &
  \multicolumn{1}{c}{C}  \\
 \midrule
 $\tilde\Xi_{11}$ & -1 & 0 & 0 & 0 & 0 & 0 & 0 & 0 & 0 & 0 & 0 & 0 & 0 & 0 & 0 & 0 & 0 & 0 \\ 
  $\tilde\Xi_{22}$ & -4 & -3 & -3 & -3 & -1 & -2 & 0 & 0 & 0 & 0 & 0 & 0 & 0 & 0 & 0 & 0 & 0 & 0 \\ 
  $\tilde\Xi_{33}$ & -1 & -1 & -1 & -1 & -1 & -2 & 1 & 0 & 0 & 1 & 0 & 0 & 0 & 0 & 0 & 0 & 0 & 0 \\ 
  $\tilde\Xi_{44}$ & -2 & -1 & -2 & -2 & -2 & -2 & 0 & 0 & 0 & 0 & 0 & 0 & 0 & 0 & 0 & 0 & 0 & 0 \\ 
  $\tilde\Xi_{55}$ & 0 & -1 & -1 & 0 & 1 & 0 & 0 & 0 & 0 & 0 & 0 & -1 & 0 & 0 & 0 & 0 & 0 & 0 \\ 
  $\tilde\Xi_{66}$ & -3 & -3 & -2 & -2 & -3 & -1 & -1 & 0 & 0 & -1 & -1 & -1 & 0 & 0 & 0 & 0 & 0 & 0 \\ 
  $\Lambda_{11}$ & -2 & -1 & -1 & -2 & -1 & -1 & 0 & 0 & 0 & 0 & 0 & 0 & 0 & 0 & 0 & 0 & 0 & 0 \\ 
  $\Lambda_{21}$ & -2 & -1 & -1 & -2 & -1 & -1 & 0 & 0 & 0 & 0 & 0 & 0 & 0 & 0 & 0 & 0 & 0 & 0 \\ 
  $\Lambda_{31}$ & -2 & -1 & -1 & -2 & -1 & -1 & 0 & 0 & 0 & 0 & 0 & 0 & 0 & 0 & 0 & 0 & 0 & 0 \\ 
  $\Lambda_{42}$ & -1 & -1 & -1 & -2 & -1 & 0 & 0 & 0 & 0 & 0 & 0 & 0 & 0 & 0 & 0 & 0 & 0 & 0 \\ 
  $\Lambda_{52}$ & -1 & -1 & -1 & -2 & -1 & 0 & 0 & 0 & 0 & 0 & 0 & 0 & 0 & 0 & 0 & 0 & 0 & 0 \\ 
  $\Lambda_{62}$ & -1 & -1 & -1 & -2 & -1 & 0 & 0 & 0 & 0 & 0 & 0 & 0 & 0 & 0 & 0 & 0 & 0 & 0 \\ 
  $\Phi_{11}$ & -6 & -4$^{a}$ & -4$^{a}$ & -7 & -4 & -3$^{a}$ & -1 & -2$^{a}$ & -2$^{a}$ & -1 & -1 & -1$^{a}$ & -1 & -2$^{a}$ & -2$^{a}$ & -1 & -1 & -2$^{a}$ \\ 
  $\Phi_{22}$ & -4 & -2$^{a}$ & -3$^{a}$ & -6 & -3 & -2$^{a}$ & -1 & -2$^{a}$ & -3$^{a}$ & -1 & -1 & -1$^{a}$ & -1 & -2$^{a}$ & -2$^{a}$ & -1 & -1 & -1$^{a}$ \\ 
  $\Phi_{12}$ & - & - & - & 16 & 4$^{a}$ & 2$^{a}$ & - & - & - & 0 & 1$^{a}$ & 1$^{a}$ &  &  &  & -1 & -2$^{a}$ & -2$^{a}$ \\ 
  $\Phi_{21}$ &  -& - & - & 9 & 4$^{a}$ & 3$^{a}$ &  -& - & - & 1 & 1$^{a}$ & 1$^{a}$ &  &  &  & 0 & -1$^{a}$ & 0$^{a}$ \\ 
  $\Gamma_{1}$ & 0 & -3 & -3$^{a}$ & - &-  & - & -1 & 0 & 0$^{a}$ &-  &  -& - & 0 & 0 & 3$^{a}$ & - &-  &  \\ 
  $\Gamma_{2}$ & 1 & 0 & 5$^{a}$ &-  & - &  -& 0 & 0 & 2$^{a}$ &  -&-  &  -& 0 & 0 & 2$^{a}$ & - & - &  \\ 
    \bottomrule
  \end{tabular}
   \begin{tablenotes}
    \item[a] Parameter generated as time-invariant but estimated as time-varying. 
   \item Note. Unless otherwise noted parameters were generated and estimated as time-invariant. Descriptions of sub-conditions A,B, and C provided in Table~\ref{tab:sim}. Cells containing a single - were generated and estimated as time-varying.
 \end{tablenotes}
  \end{threeparttable}
}
\end{table}

\subsubsection{Accuracy of Estimated Time-Varying Parameters}

\begin{figure}[H]
\caption{Mean Time-Varying Parameter Estimates for Simulation 1C}
\label{fig:sim1}
\centering
\includegraphics[width=1\textwidth]{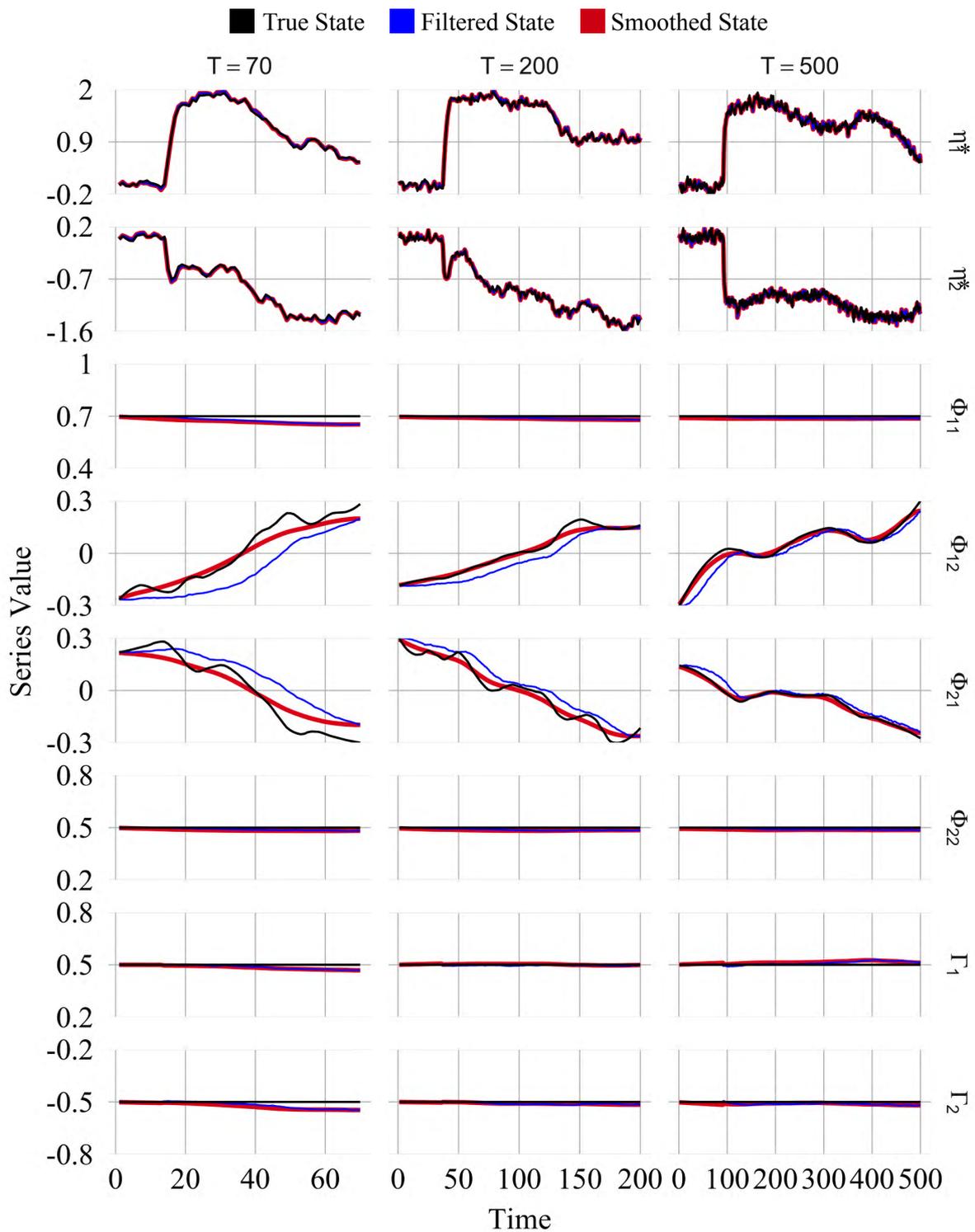}
\end{figure}

\begin{figure}[H]
\caption{Mean Time-Varying Parameter Estimates for Simulation 2C}
\label{fig:sim2}
\centering
\includegraphics[width=1\textwidth]{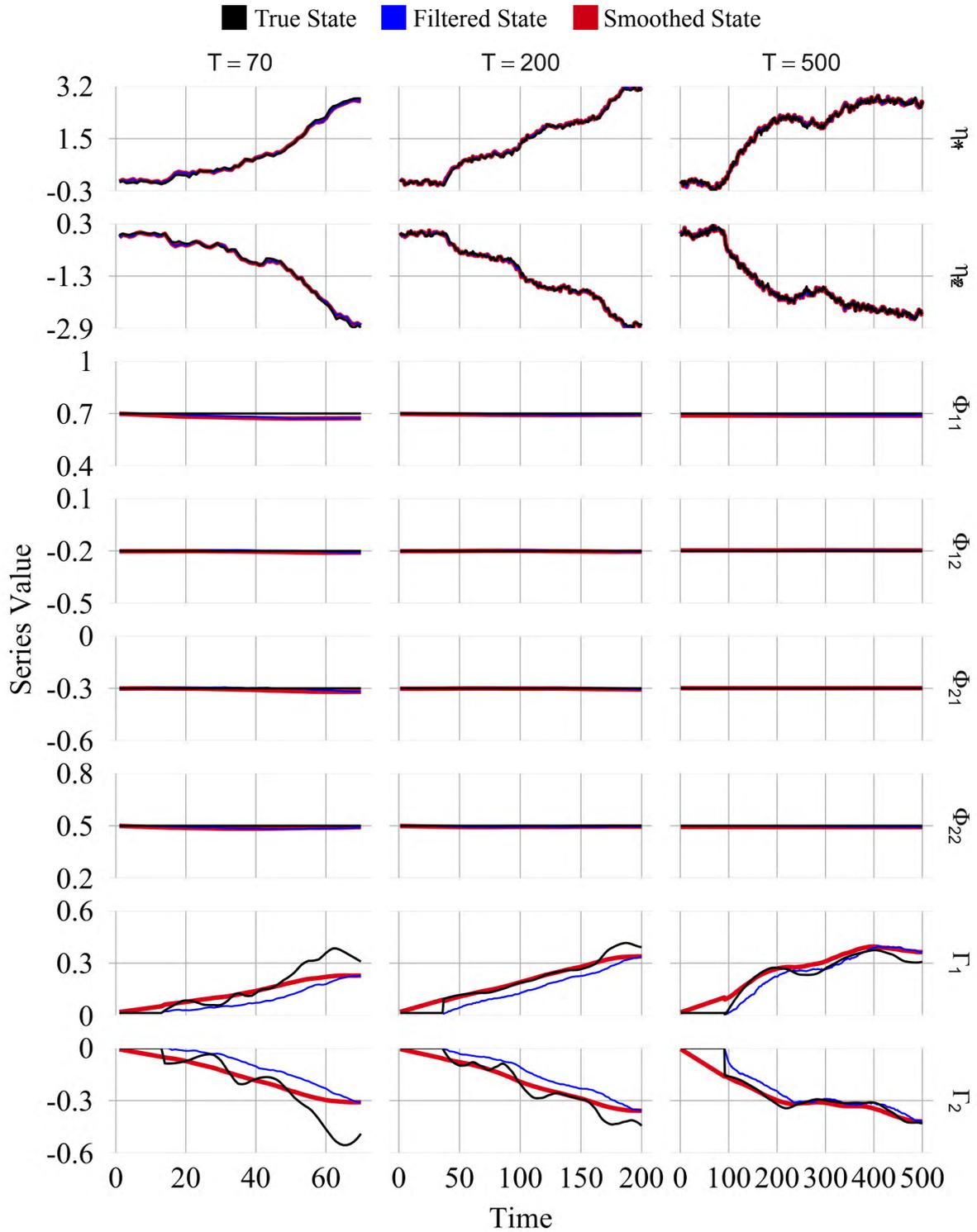}
\end{figure}

As we did not see major differences across the simulation sub-conditions we only present graphical depictions of the results from sub-condition C, where the most parameters were allowed to vary in time.  The true (or data generating) and mean estimated parameter values for all time-varying estimates in simulations 1 and 2 are provided in Figures~\ref{fig:sim1} and \ref{fig:sim2}, respectively.  First let us consider simulation 1, where only the cross-regressive parameters, $\Phi_{12}$ and $\Phi_{21}$ were generated as time-varying. In Figure~\ref{fig:sim1} it is clear the filtered and smoothed estimates of the latent states $\eta^{*}_{1}$ and $\eta^{*}_{2}$ accurately recovered the true data generating trajectories across all examined time series lengths. In addition the smoothed estimates of $\Phi_{12}$ and $\Phi_{21}$ also captured the augmented state elements well, with accuracy increasing as time series length increased.  The filtered estimates, however, tended to overestimate the magnitude of the true coefficient at the smaller time series lengths.  In terms of the parameters which were generated as time-invariant but estimated as time-varying the algorithm appears to have done well in the aggregate in terms of characterizing these parameters as constant. A similar pattern emerged for simulation 2 (see Figure~\ref{fig:sim2}), however, the smoothed estimates for the time-varying treatment effects tended to overestimate the parameter prior to the beginning of the intervention, while the filtered estimates better captured this change.

\subsubsection{Standard Deviation of Coefficient Estimates}

Efficiency of the parameter estimates across all model specifications was assessed using the standard deviation of parameter estimates within each simulation block (see Table~\ref{tab:sd}). Generally, no consistent pattern of changes in parameter variability was observed across the different simulations sub-conditions, indicating there was little impact on parameter variability when allowing additional time-varying parameters in the estimated model, even when those parameters were in fact generated as time-invariant. As was expected across all sub-conditions variability decreased as the time series length increased. 

\begin{table}[ht]
\centering
\caption{Standard Deviation of Parameters Generated as Time-Invariant }
\label{tab:sd} 
\scalebox{0.7}{
  \begin{threeparttable}
\begin{tabular}{rrrrrrrrrrrrrrrrrrr}
  \toprule
  &   \multicolumn{18}{c}{Time Series Length}  \\
 \cmidrule(lr){2-19} & \multicolumn{6}{c}{T = 70} &
  \multicolumn{6}{c}{T = 200} &
  \multicolumn{6}{c}{T = 500} \\
 \cmidrule(lr){2-7} \cmidrule(lr){8-13}\cmidrule(lr){14-19} & \multicolumn{6}{c}{Simulation Condition} &
  \multicolumn{6}{c}{Simulation Condition} &
  \multicolumn{6}{c}{Simulation Condition} \\
 \cmidrule(lr){2-7} \cmidrule(lr){8-13}\cmidrule(lr){14-19} &  \multicolumn{3}{c}{1} &
  \multicolumn{3}{c}{2} &
  \multicolumn{3}{c}{1} &
  \multicolumn{3}{c}{2} &
  \multicolumn{3}{c}{1} &
  \multicolumn{3}{c}{2} \\
 \cmidrule(lr){2-4} \cmidrule(lr){5-7}\cmidrule(lr){8-10}
  \cmidrule(lr){11-13} \cmidrule(lr){14-16}\cmidrule(lr){17-19} \multicolumn{1}{c}{Parameter} & 
  \multicolumn{1}{c}{A} &
  \multicolumn{1}{c}{B} &
  \multicolumn{1}{c}{C} &
  \multicolumn{1}{c}{A} &
  \multicolumn{1}{c}{B} &
  \multicolumn{1}{c}{C} &
  \multicolumn{1}{c}{A} &
  \multicolumn{1}{c}{B} &
  \multicolumn{1}{c}{C} &
  \multicolumn{1}{c}{A} &
  \multicolumn{1}{c}{B} &
  \multicolumn{1}{c}{C} &
  \multicolumn{1}{c}{A} &
  \multicolumn{1}{c}{B} &
  \multicolumn{1}{c}{C} &
  \multicolumn{1}{c}{A} &
  \multicolumn{1}{c}{B} &
  \multicolumn{1}{c}{C}  \\
 \midrule
$\tilde\Xi_{11}$ & 0.05 & 0.05 & 0.04 & 0.05 & 0.05 & 0.05 & 0.02 & 0.01 & 0.01 & 0.02 & 0.02 & 0.02 & 0.00 & 0.00 & 0.00 & 0.01 & 0.00 & 0.00 \\ 
  $\tilde\Xi_{22}$ & 0.05 & 0.05 & 0.04 & 0.05 & 0.05 & 0.05 & 0.02 & 0.01 & 0.01 & 0.02 & 0.02 & 0.02 & 0.00 & 0.00 & 0.00 & 0.01 & 0.00 & 0.01 \\ 
  $\tilde\Xi_{33}$ & 0.05 & 0.05 & 0.05 & 0.05 & 0.05 & 0.05 & 0.02 & 0.01 & 0.01 & 0.02 & 0.02 & 0.02 & 0.00 & 0.00 & 0.00 & 0.00 & 0.00 & 0.00 \\ 
  $\tilde\Xi_{44}$ & 0.05 & 0.04 & 0.04 & 0.05 & 0.05 & 0.04 & 0.01 & 0.01 & 0.01 & 0.02 & 0.02 & 0.02 & 0.00 & 0.00 & 0.00 & 0.01 & 0.00 & 0.00 \\ 
  $\tilde\Xi_{55}$ & 0.05 & 0.05 & 0.05 & 0.05 & 0.05 & 0.05 & 0.01 & 0.01 & 0.01 & 0.02 & 0.02 & 0.02 & 0.00 & 0.00 & 0.00 & 0.00 & 0.00 & 0.00 \\ 
  $\tilde\Xi_{66}$ & 0.05 & 0.05 & 0.05 & 0.05 & 0.05 & 0.05 & 0.01 & 0.01 & 0.01 & 0.02 & 0.02 & 0.02 & 0.00 & 0.00 & 0.00 & 0.00 & 0.00 & 0.00 \\ 
  $\Lambda_{11}$ & 0.08 & 0.07 & 0.06 & 0.10 & 0.08 & 0.07 & 0.02 & 0.02 & 0.02 & 0.02 & 0.02 & 0.02 & 0.01 & 0.01 & 0.01 & 0.01 & 0.01 & 0.01 \\ 
  $\Lambda_{21}$ & 0.08 & 0.07 & 0.06 & 0.10 & 0.08 & 0.07 & 0.02 & 0.02 & 0.02 & 0.02 & 0.02 & 0.02 & 0.01 & 0.01 & 0.01 & 0.01 & 0.01 & 0.01 \\ 
  $\Lambda_{31}$ & 0.08 & 0.07 & 0.06 & 0.10 & 0.08 & 0.07 & 0.02 & 0.02 & 0.02 & 0.02 & 0.02 & 0.02 & 0.01 & 0.01 & 0.01 & 0.01 & 0.01 & 0.01 \\ 
  $\Lambda_{42}$ & 0.07 & 0.06 & 0.06 & 0.09 & 0.08 & 0.07 & 0.02 & 0.02 & 0.02 & 0.02 & 0.02 & 0.02 & 0.01 & 0.01 & 0.01 & 0.01 & 0.01 & 0.01 \\ 
  $\Lambda_{52}$ & 0.07 & 0.07 & 0.06 & 0.09 & 0.08 & 0.07 & 0.02 & 0.02 & 0.02 & 0.02 & 0.02 & 0.02 & 0.01 & 0.01 & 0.01 & 0.01 & 0.01 & 0.01 \\ 
  $\Lambda_{62}$ & 0.07 & 0.07 & 0.06 & 0.09 & 0.08 & 0.07 & 0.02 & 0.02 & 0.02 & 0.02 & 0.02 & 0.02 & 0.01 & 0.01 & 0.01 & 0.01 & 0.01 & 0.01 \\ 
  $\Phi_{11}$ & 0.08 & - & - & 0.12 & 0.10 & - & 0.03 & - & - & 0.03 & 0.02 & - & 0.01 & - & - & 0.01 & 0.01 & - \\ 
  $\Phi_{22}$ & 0.09 & - & - & 0.13 & 0.10 & - & 0.02 & - & - & 0.02 & 0.03 & - & 0.01 & - & - & 0.01 & 0.01 &- \\ 
  $\Phi_{12}$ & - & - & - & 0.11 & - & - &  -&-  & - & 0.03 & - & - &  -& - & - & 0.01 & - & - \\ 
  $\Phi_{21}$ & - & - &  -& 0.10 & - & - &  -&  -& -& 0.02 & - & - &-  & - & - & 0.01 & - & - \\ 
  $\Gamma_{1}$ & 0.15 & 0.10 & - &-  & - &-  & 0.01 & 0.01 & - &  -& - & - & 0.01 & 0.00 & - & - & - &-  \\ 
  $\Gamma_{2}$ & 0.12 & 0.11 & -& - &  -& - & 0.02 & 0.02 & - & - &  -& - & 0.01 & 0.00 & - & - &  -&  -\\ 
  \bottomrule
  \end{tabular}
     \begin{tablenotes}
       \item Note. Unless otherwise noted parameters were generated and estimated as time-invariant. Descriptions of sub-conditions A,B, and C provided in Table~\ref{tab:sim}. Cells containing a single - were estimated as time-varying. Standard deviations were rounded to the third decimal place.
 \end{tablenotes}
  \end{threeparttable}
}
\end{table}

\subsubsection{Detection of Time-Invariant and Time-Varying Parameters}

Finally we consider the accuracy of the smoothed parameter estimates in determining whether a parameter was is time-varying or time-invariant. As mentioned previously this is a conservative criteria as we may reasonably expect the smoothed estimations of a time-invariant parameter to have some small amount of variation around the true value.  However, as the purpose of this study was not the evaluation of secondary procedures for determining time-invariance we evaluated the smoothed estimates for this purpose.  Smoothed parameter estimates were considered to be time-varying if those estimates had some non-zero variance across the time series length, and time-invariant if the smoothed parameter estimates were constant (with zero variance). 

As can be seen from Table~\ref{tab:class} the true-time varying parameters were classified as such $100\%$ of the time, even at the smaller sample sizes. Except for the constant treatment effects in simulation 1 the time-invariant parameters were also correctly classified as time-invariant between $80\%-99\%$ of the time, with classification improving at the larger sample sizes.  Unlike the other time-invariant parameters the smoothed estimates of the treatment effects ($\Gamma_{1},\Gamma_{2}$) showed some nonzero variance in a larger proportion of replications. However, as can be seen from   Figure~\ref{fig:sim1} where the means of the smoothed estimates of $\Gamma_{1}$ and $\Gamma_{2}$ track tightly with the constant data generating values and the small variances for those parameters in Condition 1C (seeTable~\ref{tab:sd}) one would likely classify $\Gamma_{1}$ and $\Gamma_{2}$ as time-invariant based on a simple plot of the parameter estimates across time. 

The results for the convenience sample taken here indicate a complex pattern of dependencies among the parameters and constructs of interest across time.  Visually, subjects 1, 2, 5, 6, and 7 all show an increasing impact of the treatment on positive affect over time.  This, however, does not always translate to concurrent increases in the levels of positive affect, as these changes often coincide with changes in other model parameters. As we did not let the intervention directly impact the parameters themselves, although this is certainly possible using the augmented state vector in \ref{ssn.eqn}, it is difficult to ascertain whether the reorganization of dynamics occurring in $\boldsymbol{\Phi}$ reflects a re-organization of the system that is inherently \emph{self-organizing} or a result of external influence. Although the simulation results presented here suggest one is unlikely to observe broad variability in parameter trajectories (as is evident among many of the subjects here) if the parameter is in fact time-invariant, future work should examine the recovery of more complex patterns of parameter change to allow for more confident conclusions to be drawn.  The results from this empirical example also point to the utility of looking at the impact of exogenous covariates not only on constructs themselves, but on parameters that may coincide with substantively interesting aspects of the theory governing model construction.  

\begin{table}[ht]
\centering
\caption{Percentage of Accurately Classified Time-Varying and Time-Invariant Parameters Based on Smoothed Estimates} 
\label{tab:class}
\scalebox{0.65}{
  \begin{threeparttable}
\begin{tabular}{rrrrrrrrrrrrrrrrrrr}
  \toprule
  &   \multicolumn{18}{c}{Time Series Length}  \\
 \cmidrule(lr){2-19} & \multicolumn{6}{c}{T = 70} &
  \multicolumn{6}{c}{T = 200} &
  \multicolumn{6}{c}{T = 500} \\
 \cmidrule(lr){2-7} \cmidrule(lr){8-13}\cmidrule(lr){14-19} & \multicolumn{6}{c}{Simulation Condition} &
  \multicolumn{6}{c}{Simulation Condition} &
  \multicolumn{6}{c}{Simulation Condition} \\
 \cmidrule(lr){2-7} \cmidrule(lr){8-13}\cmidrule(lr){14-19} &  \multicolumn{3}{c}{1} &
  \multicolumn{3}{c}{2} &
  \multicolumn{3}{c}{1} &
  \multicolumn{3}{c}{2} &
  \multicolumn{3}{c}{1} &
  \multicolumn{3}{c}{2} \\
 \cmidrule(lr){2-4} \cmidrule(lr){5-7}\cmidrule(lr){8-10}
  \cmidrule(lr){11-13} \cmidrule(lr){14-16}\cmidrule(lr){17-19} \multicolumn{1}{c}{Parameter} & 
  \multicolumn{1}{c}{A} &
  \multicolumn{1}{c}{B} &
  \multicolumn{1}{c}{C} &
  \multicolumn{1}{c}{A} &
  \multicolumn{1}{c}{B} &
  \multicolumn{1}{c}{C} &
  \multicolumn{1}{c}{A} &
  \multicolumn{1}{c}{B} &
  \multicolumn{1}{c}{C} &
  \multicolumn{1}{c}{A} &
  \multicolumn{1}{c}{B} &
  \multicolumn{1}{c}{C} &
  \multicolumn{1}{c}{A} &
  \multicolumn{1}{c}{B} &
  \multicolumn{1}{c}{C} &
  \multicolumn{1}{c}{A} &
  \multicolumn{1}{c}{B} &
  \multicolumn{1}{c}{C}  \\
  \midrule
$\Phi_{12}$ & 100$^a$ & 100$^a$ & 100$^a$ & -& 85 & 87 & 100$^a$ & 100$^a$ & 100$^a$ & - & 93 & 94 & 100$^a$ & 100$^a$ & 100$^a$ &-  & 99 & 99 \\ 
  $\Phi_{21}$ & 100$^a$ & 100$^a$ & 100$^a$ & - & 80 & 83 & 100$^a$ & 100$^a$ & 100$^a$ & - & 93 & 96 & 100$^a$ & 100$^a$ & 100$^a$ & - & 99 & 100 \\ 
  $\Phi_{11}$ & - & 80 & 78 &  -& - & 87 & - & 86 & 87 &-  &  -& 96 &-  & 88 & 93 &-  &-  & 99 \\ 
  $\Phi_{22}$ & - & 85 & 87 & - &  -& 85 &  & 81 & 86 &  -& - & 94 & - & 80 & 92 & - &  -& 99 \\ 
  $\Gamma_{1}$ & - & - & 17 & 100$^a$ & 100$^a$ & 100$^a$ &  -&  -& 15 & 100$^a$ & 100$^a$ & 100$^a$ & - &  -& 15 & 100$^a$ & 100$^a$ & 100$^a$ \\ 
  $\Gamma_{2}$ &-  &  -& 19 & 100$^a$ & 100$^a$ & 100$^a$ & - &  -& 20 & 100$^a$ & 100$^a$ & 100$^a$ &  -&  -& 16 & 100$^a$ & 100$^a$ & 100$^a$ \\ 
  \bottomrule
  \end{tabular}
    \begin{tablenotes}
    \item[a] Parameter generated as time-varying. All other parameters were generated as time-invariant.   \item Note. Cells containing a single - were generated and estimated as time-invariant and for this reason misclassification was not possible.
 \end{tablenotes}
  \end{threeparttable}

}
\end{table}

\textcolor{black}{\subsubsection{Comparison of Filter Implementations}}

\textcolor{black}{As expected the square-root version of the second-order EKF provided a number of benefits when compared to the standard SEKF implementation. As we had no reason to believe our results would differ across the two simulations we only compared to two approaches for Simulation 1.  The mean relative bias and standard deviation of the parameter estimates obtained from the SEKF are presented in Table \ref{tab:sekf}. In terms of relative bias both approaches performed well, although in aggregate the SR-SEKF obtained lower relative bias across all parameter types for the models considered here. This difference was most pronounced for the structural model parameters (or dynamics) at the smallest sample size.  The estimates obtained from SR-SEKF also exhibited less variability, consistent with the notion that square-root filters can reduce the propagation of numerical error across iterations. }

\textcolor{black}{We also hypothesized the SR-SEKF would bring additional computational benefits when compared to the standard SEKF algorithm.  To assess this we recorded the mean number of iterations and the percentage of converged datasets per condition for Simulation 1. These outcome measures can be found in Table \ref{tab:comp}. Consistent with general results in the KF literature, in aggregate the SR-SEKF required fewer iterations per condition and exhibited a modest increase in the percentage of converged datasets.}

\begin{table}[H]
\centering
\caption{Mean Percentage of Relative Bias and Standard Deviation for the SEKF} 
\label{tab:sekf}
\scalebox{0.73}{
\begin{threeparttable}
\begin{tabular}{rrrrrrrrrrrrrrrrrrr}
  \toprule
  &   \multicolumn{18}{c}{Outcome Measure}  \\
 \cmidrule(lr){2-19} & \multicolumn{9}{c}{Relative Bias} &
  \multicolumn{9}{c}{Standard Deviation} \\
 \cmidrule(lr){2-10}\cmidrule(lr){11-19} & \multicolumn{3}{c}{T = 70} &
  \multicolumn{3}{c}{T = 200} &
  \multicolumn{3}{c}{T = 500} &
  \multicolumn{3}{c}{T = 70} &
  \multicolumn{3}{c}{T = 200} &
  \multicolumn{3}{c}{T = 500} \\
 \cmidrule(lr){2-4} \cmidrule(lr){5-7}\cmidrule(lr){8-10}
  \cmidrule(lr){11-13} \cmidrule(lr){14-16}\cmidrule(lr){17-19} & \multicolumn{3}{c}{Condition} &
  \multicolumn{3}{c}{Condition} &
  \multicolumn{3}{c}{Condition} &
  \multicolumn{3}{c}{Condition} &
  \multicolumn{3}{c}{Condition} &
  \multicolumn{3}{c}{Condition} \\
 \cmidrule(lr){2-4} \cmidrule(lr){5-7}\cmidrule(lr){8-10}
  \cmidrule(lr){11-13} \cmidrule(lr){14-16}\cmidrule(lr){17-19} \multicolumn{1}{c}{Parameter} & 
  \multicolumn{1}{c}{1A} &
  \multicolumn{1}{c}{1B} &
  \multicolumn{1}{c}{1C} &
  \multicolumn{1}{c}{1A} &
  \multicolumn{1}{c}{1B} &
  \multicolumn{1}{c}{1C} &
  \multicolumn{1}{c}{1A} &
  \multicolumn{1}{c}{1B} &
  \multicolumn{1}{c}{1C} &
  \multicolumn{1}{c}{1A} &
  \multicolumn{1}{c}{1B} &
  \multicolumn{1}{c}{1C} &
  \multicolumn{1}{c}{1A} &
  \multicolumn{1}{c}{1B} &
  \multicolumn{1}{c}{1C} &
  \multicolumn{1}{c}{1A} &
  \multicolumn{1}{c}{1B} &
  \multicolumn{1}{c}{1C}  \\
 \midrule
$\tilde\Xi_{11}$ & -1 & -1 & 0 & 0 & -0 & 0 & 0 & 0 & 0 & 0.04 & 0.04 & 0.04 & 0.02 & 0.02 & 0.02 & 0.01 & 0.01 & 0.01 \\ 
  $\tilde\Xi_{22}$ & -3 & -4 & -4 & 1 & 1 & 1 & 0 & 0 & 0 & 0.05 & 0.04 & 0.04 & 0.02 & 0.02 & 0.02 & 0.01 & 0.01 & 0.01 \\ 
  $\tilde\Xi_{33}$ & -2 & -2 & -2 & 0 & 1 & 0 & 0 & 0 & 0 & 0.05 & 0.04 & 0.04 & 0.02 & 0.02 & 0.02 & 0.01 & 0.01 & 0.01 \\ 
  $\tilde\Xi_{44}$ & -3 & -2 & -3 & 0 & 0 & 0 & 0 & -1 & 0 & 0.05 & 0.04 & 0.04 & 0.02 & 0.02 & 0.02 & 0.01 & 0.01 & 0.01 \\ 
  $\tilde\Xi_{55}$ & 0 & 1 & 1 & 0 & 0 & 1 & 0 & 0 & 0 & 0.05 & 0.04 & 0.04 & 0.02 & 0.02 & 0.02 & 0.01 & 0.01 & 0.01 \\ 
  $\tilde\Xi_{66}$ & -2 & -2 & -1 & 0 & 0 & -1 & 0 & 1 & 0 & 0.05 & 0.05 & 0.04 & 0.02 & 0.02 & 0.02 & 0.01 & 0.01 & 0.01 \\ 
  $\Lambda_{11}$ & -2 & -1 & -1 & -1 & 0 & 0 & 0 & 0 & 0 & 0.09 & 0.08 & 0.07 & 0.04 & 0.03 & 0.03 & 0.01 & 0.01 & 0.01 \\ 
  $\Lambda_{21}$ & -2 & -1 & -1 & -1 & 0 & 0 & 0 & 0 & 0 & 0.09 & 0.08 & 0.07 & 0.04 & 0.03 & 0.03 & 0.01 & 0.01 & 0.01 \\ 
  $\Lambda_{31}$ & -1 & -1 & -1 & 0 & -1 & 0 & 0 & 0 & 0 & 0.09 & 0.08 & 0.07 & 0.04 & 0.03 & 0.03 & 0.01 & 0.01 & 0.01 \\ 
  $\Lambda_{42}$ & -2 & -1 & -1 & -1 & 0 & 0 & 0 & 0 & 0 & 0.08 & 0.07 & 0.07 & 0.04 & 0.03 & 0.03 & 0.02 & 0.01 & 0.01 \\ 
  $\Lambda_{52}$ & -2 & -1 & -1 & -1 & -1 & -1 & 0 & 0 & 0 & 0.08 & 0.07 & 0.07 & 0.04 & 0.03 & 0.03 & 0.01 & 0.01 & 0.01 \\ 
  $\Lambda_{62}$ & -2 & -1 & -1 & -1 & -1 & -0 & -0 & -0 & -0 & 0.08 & 0.07 & 0.07 & 0.04 & 0.03 & 0.04 & 0.01 & 0.01 & 0.01 \\ 
  $\Phi_{11}$ & -4 & -4$^{a}$ & -5$^{a}$ & -1 & -2$^{a}$ & -2$^{a}$ & -1 & -1$^{a}$ & -2$^{a}$ & 0.08 & - & - & 0.04 & - &  -& 0.02 & - & - \\ 
  $\Phi_{22}$ & -6 & -2$^{a}$ & -3$^{a}$ & -2 & -1$^{a}$ & -2$^{a}$ & -1 & -1$^{a}$ & -1$^{a}$ & 0.10 & - & - & 0.04 & - & - & 0.01 & - & - \\ 
  $\Gamma_{1}$ & -4 & -4 & -2$^{a}$ & -1 & -1 & -1$^{a}$ & 0 & 0 & 1$^{a}$ & 0.17 & 0.14 & - & 0.08 & 0.06 & - & 0.02 & 0.02 & - \\ 
  $\Gamma_{2}$ & 10 & 4 & 5$^{a}$ & 3 & 2 & 2$^{a}$ & 1 & 1 & 2$^{a}$ & 0.17 & 0.13 & - & 0.07 & 0.05 & - & 0.02 & 0.02 & - \\ 
  \bottomrule
  \end{tabular}
     \begin{tablenotes}
    \item[a] Parameter generated as time-invariant but estimated as time-varying. 
   \item Note. Unless otherwise noted parameters were generated and estimated as time-invariant. Descriptions of sub-conditions A,B, and C provided in Table~\ref{tab:sim}. Cells containing a single - were estimated as time-varying.
 \end{tablenotes}
  \end{threeparttable}
}
\end{table}

\begin{table}[ht]
\centering
\caption{Comparison of Computational Efficiency and Convergence Rates for EKF Implementations} 
\label{tab:comp}
\scalebox{0.7}{
\begin{tabular}{lrrrrrrrrrrrrrrrrrr}
  \toprule
  &   \multicolumn{18}{c}{Method}  \\
 \cmidrule(lr){2-19} & \multicolumn{9}{c}{Second-Order EKF} &
  \multicolumn{9}{c}{Square-Root Second-Order EKF} \\
 \cmidrule(lr){2-10}\cmidrule(lr){11-19} & \multicolumn{3}{c}{T = 70} &
  \multicolumn{3}{c}{T = 200} &
  \multicolumn{3}{c}{T = 500} &
  \multicolumn{3}{c}{T = 70} &
  \multicolumn{3}{c}{T = 200} &
  \multicolumn{3}{c}{T = 500} \\
 \cmidrule(lr){2-4} \cmidrule(lr){5-7}\cmidrule(lr){8-10}
  \cmidrule(lr){11-13} \cmidrule(lr){14-16}\cmidrule(lr){17-19} & \multicolumn{3}{c}{Condition} &
  \multicolumn{3}{c}{Condition} &
  \multicolumn{3}{c}{Condition} &
  \multicolumn{3}{c}{Condition} &
  \multicolumn{3}{c}{Condition} &
  \multicolumn{3}{c}{Condition} \\
 \cmidrule(lr){2-4} \cmidrule(lr){5-7}\cmidrule(lr){8-10}
  \cmidrule(lr){11-13} \cmidrule(lr){14-16}\cmidrule(lr){17-19} \multicolumn{1}{c}{Measure} & 
  \multicolumn{1}{c}{A} &
  \multicolumn{1}{c}{B} &
  \multicolumn{1}{c}{C} &
  \multicolumn{1}{c}{A} &
  \multicolumn{1}{c}{B} &
  \multicolumn{1}{c}{C} &
  \multicolumn{1}{c}{A} &
  \multicolumn{1}{c}{B} &
  \multicolumn{1}{c}{C} &
  \multicolumn{1}{c}{A} &
  \multicolumn{1}{c}{B} &
  \multicolumn{1}{c}{C} &
  \multicolumn{1}{c}{A} &
  \multicolumn{1}{c}{B} &
  \multicolumn{1}{c}{C} &
  \multicolumn{1}{c}{A} &
  \multicolumn{1}{c}{B} &
  \multicolumn{1}{c}{C}  \\
 \midrule
Mean Number of Iterations & 63 & 57 & 54 & 40 & 41 & 41 & 30 & 26 & 27 & 57 & 58 & 61 & 40 & 36 & 35 & 27 & 24 & 26 \\ 
$\%$ of Converged Datasets & 94 & 96 & 93 & 99 & 100 & 98 & 99 & 100 & 99 & 96 & 95 & 95 & 100 & 100 & 100 & 100 & 100 & 100 \\ 
\bottomrule
  \end{tabular}
}
\end{table}

\section{An Empirical Example}

\begin{figure}[H]
\caption{Example Time Course of Study}
\label{fig:emp1}
\centering
\includegraphics[width=1\textwidth]{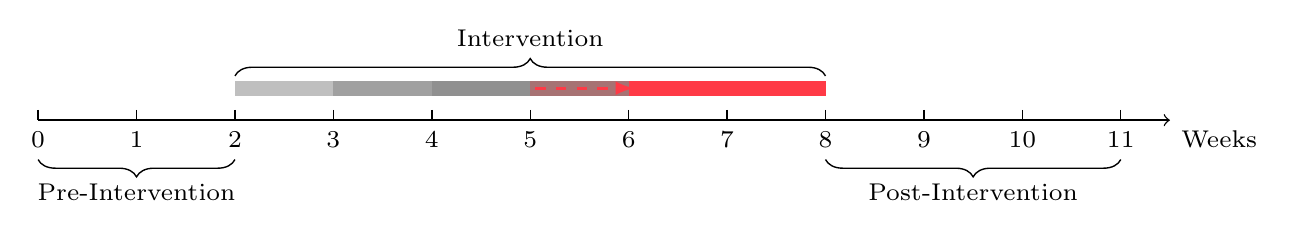}
\end{figure}

We now present an empirical example based on \citet{fredrickson2017} who examined the impact of meditation practice on the day-to-day emotional experiences of a nonclinical adult sample across an eleven week period. The first two weeks of the data collection period represented a baseline period, followed by six weeks where subjects were involved in a workshop to learn and integrate meditation practice into their daily lives. Following the termination of the intervention data was collected for an additional three weeks. A timeline of the typical intervention time course is given in Figure~\ref{fig:emp1}.  Each evening participants recorded their experience of daily emotions using the modified Differential Emotions Scale (mDES) \citep{fredrickson2013}. The mDES is a 20-item measure containing ten positive emotions and ten negative emotions. \textcolor{black}{For completeness, the full mDES questionnaire is reproduced in the Appendix. For the purpose of our current study we included all 10 indicators for each of the emotion constructs.}
 
Although missing data can be handled well within the state space framework the implementation of the SEKF used here was not designed to handle missing data. For this reason we retained subjects with less than $5\%$ missing data and imputed the missing values univariately using the predicted values from a single run of the Kalman Filter. This procedure left us with seven subjects on which to conduct our analysis.  For each of these individuals we fit a two factor (positive and negative affect) nonlinear state space model where the autoregressive coefficients, cross-regressive coefficients and treatment effects were all allowed to vary across time.  Starting values for each subject were obtained by estimating a time-invariant model using the pseudo-ML estimator.  The binary exogenous input was used to indicate whether each timepoint occurred prior to or following the onset of the meditation intervention. Results for the dynamic parameters for each subjects are provided in Figure~\ref{fig:emp2} where the vertical dashed lines indicate the onset of the intervention.

In Figure~\ref{fig:emp2} the first row gives the trajectory of positive ($\eta^{*}_{1}$) and negative affect ($\eta^{*}_{2}$) across the observation period. In this row we should see the impact of the intervention on the two constructs of interest. Large oscillations in $\eta^{*}_{1}$ and $\eta^{*}_{2}$ as in the negative affect of Subject 2 indicate periodic swings that may be related to weekly patterns such as a weekend effect. The second row illustrates the potentially time-varying trajectories of the autoregressive coefficients for positive and negative affect. The closer the autoregressive coefficient gets to unity the more persistent the series becomes, meaning the value of the series at its previous time point better and better predicts its current value.  A negative autoregressive coefficient (as in Subject 3) indicates a type of oscillation where an observation is likely to be above average if its previous value was below average, and vice versa. The third row details the cross-regressive coefficients, or the lagged influence of negative affect on positive affect ($\Phi_{12}$) and positive affect on negative affect ($\Phi_{21}$). Changes in these coefficient over time indicates increasing or decreasing influence of positive and negative affect on one another. Finally, row four provides time series for the continuing effects of the intervention on positive ($\Gamma_{1}$) and negative affect ($\Gamma_{2}$) across the observation period. 

\begin{landscape}
\begin{figure}[h]
\caption{Results for 7 Subjects}
\label{fig:emp2}
\centering
\includegraphics[width=1.2\textwidth]{emp1.pdf}
\caption*{Note. Horizontal dashed line indicates the onset of the intervention. $\eta^{*}_{1}$ is positive affect and  $\eta^{*}_{2}$ is negative affect.}
\end{figure}
\end{landscape}

\section{Discussion}

In this paper we introduced a nonlinear state space model and associated estimation routines based on a square-root second-order Extended Kalman Filter (SR-SEKF). We demonstrated how the state vector of a linear state space model could be augmented to include arbitrarily time-varying parameters when little information about the parametric form of change is available.  Through a series of simulation studies we showed that the described algorithm accurately recovers the unobserved states in the case of a bivariate dynamic factor model with time-varying dynamics. We found the point estimates for the time-invariant parameters to exhibit low levels of bias even for the shortest time series lengths of $T=70$.  For the case of two time-varying parameters we found that increasing the number of theoretically time-varying parameters in the estimated model has very little impact on the accuracy or variability of the estimated parameters. Furthermore, the smoothed estimates of the time-varying parameters themselves provide a reasonably accurate index of the true parameter variability for the models under study. \textcolor{black}{Despite this performance one must always exercise caution when fitting nonlinear models to shorter time series lengths as real-world data is likely to contain more structural specifications than those considered in our synthetic data examples.}

In addition to the novel insights gained though this work a number of limitations are also worth discussing. First and foremost we did not evaluate the standard errors generated as a by-product of the likelihood maximization described above.  Theoretically, the observed or expected information matrices produced by this procedure could be used to assess the variability of the time-invariant parameter estimates. Alternative procedures for obtaining the standard errors, such as the nonparametric bootstrap, are available and future work should better examine these different options. 

Furthermore, significance tests based on the estimated error variances arising from the tuning procedure could provide additional insight into whether a parameter is truly time-varying. Relatedly, we did not explore any post-hoc metrics for determining whether a parameter estimated as time-varying does in fact vary across time.  Although it appears that the smoothed parameter estimates themselves and plots of the smoothed estimates across time does provide valuable insights on the time-varyingness of parameters, inferential tests could provide more sensitive diagnostics.  Future work should also explore which methods or tests are best suited for determining whether a parameter varies meaningfully across time. 

Second, in this paper we examined in some detail the case where multiple time-varying parameters were estimated, however, only a small subset of those estimated parameters were in fact time-varying. It is important to also explore the case where a great diversity and number of parameter types are simultaneously time-varying. This will likely introduce a number of complicating issues related to both model interpretation and parameter identification.  

We also did not explore the important concept of stochastic observability for time-varying nonlinear state space models.  Intuitively, a system is observable if one can infer its states from its output. More specifically, observability describes the possibility of inferring the current state from present and future measurements of the system. \citet{kalman1960a, kalman1960b} was the first to introduce the concept of observability for linear time-invariant systems, however, extensions to the nonlinear stochastic case are currently an active area of research. Tests for stochastic observability will likely provide novel insights into how many, and which types, of parameters can feasibly vary given a specific set of model constraints. Future work should better detail the implications of stochastic observability for the types of models most useful to psychological researchers. 

Nonlinear state space models with time-varying parameters provide a natural modeling framework from which psychological researchers can characterize and test complex theories of psychological change. Although applied researchers are increasingly collecting the types of time series data (e.g. intensive longitudinal data) appropriate for this and similar modeling approaches there is still a dearth of practical information on how these methods perform in situations relevant to applied psychological researchers. In this vein we hope our exposition of time-varying parameter state space models and their estimation, along with our empirical evaluations, contribute to bridging this gap.

\section{Appendix}

Below we have reproduced the modified Differential Emotions Scale (mDES) from \citet[][p. 45]{fredrickson2013}.\\

\noindent \textbf{Instructions}: Please think back to how you have felt during the past 24 hours. Using the 0-4 scale below, indicate the greatest amount that you have experienced each of the following feelings where 0 = not at all; 1 = A little bit; 2 = Moderately, 3 = Quite a bit, and 4 = Extremely. 

\begin{enumerate}[nosep]
\item What is the most amused, fun-loving, or silly you felt?  
\item What is the most angry, irritated, or annoyed you felt?  
\item What is the most ashamed, humiliated, or disgraced you felt?  
\item What is the most awe, wonder, or amazement you felt?  
\item What is the most contemptuous, scornful, or disdainful you felt?  
\item What is the most disgust, distaste, or revulsion you felt?  
\item What is the most embarrassed, self-conscious, or blushing you felt?  
\item What is the most grateful, appreciative, or thankful you felt?  
\item What is the most guilty, repentant, or blameworthy you felt?  
\item What is the most hate, distrust, or suspicion you felt?  
\item What is the most hopeful, optimistic, or encouraged you felt?  
\item What is the most inspired, uplifted, or elevated you felt?  
\item What is the most interested, alert, or curious you felt?  
\item What is the most joyful, glad, or happy you felt?  
\item What is the most love, closeness, or trust you felt?  
\item What is the most proud, confident, or self-assured you felt?  
\item What is the most sad, downhearted, or unhappy you felt?  
\item What is the most scared, fearful, or afraid you felt?  
\item What is the most serene, content, or peaceful you felt?  
\item What is the most stressed, nervous, or overwhelmed you felt?  
\end{enumerate}
\bibliography{references.bib}
\end{document}